\documentclass[aps,prd,preprint,groupedaddress,showpacs,preprintnumbers,amsmath,amssymb]{revtex4-1}

\usepackage{graphicx}
\usepackage{amsmath}
\usepackage{verbatim}
\usepackage{color}
\usepackage{subfigure}
\usepackage{hyperref}

\begin{document}

\preprint{SLAC-PUB-15129}

\title{QCD Analysis of the Scale-Invariance of Jets}

\author{Andrew J.~Larkoski}
\email[]{larkoski@stanford.edu}

\affiliation{SLAC National Accelerator Laboratory, Menlo Park, CA 94025}

\date{\today}

\begin{abstract}
Studying the substructure of jets has become a powerful tool for event discrimination and for studying QCD.  
Typically, jet substructure studies rely on Monte Carlo simulation for vetting their usefulness; however, when possible, it is also important to compute 
observables with analytic methods.  Here, we present a global next-to-leading-log resummation of the angular correlation function which measures the
contribution to the mass of a jet from constituents that are within an angle $R$ with respect to one another.  
For a scale-invariant jet, the angular correlation function should scale as a power of $R$.  Deviations from this behavior can be traced to
the breaking of scale invariance in QCD.
To do the resummation, we use soft-collinear
effective theory relying on the recent proof of factorization of jet observables at $e^+ e^-$ colliders. 
 Non-trivial requirements of factorization of the angular correlation function
are discussed. The calculation is compared to Monte Carlo parton shower and next-to-leading order results.  The different calculations
are important in distinct phase space regions
 and exhibit that jets in QCD are, to very good approximation, scale invariant over a wide dynamical range.
\end{abstract}


\maketitle

\section{Introduction}

The current success of the Large Hadron Collider (LHC), its high center of mass energies, its significant delivered integrated luminosity
and its high-precision experiments has ushered in a new era of particle physics.  Particles and jets with
significant transverse boosts are now being copiously produced.  An entire field of studying the
substructure of highly boosted jets has grown up out of the study of these objects and many methods have been proposed to study QCD.  In addition, procedures
for discriminating
QCD jets from jets initiated by heavy particle decays have been introduced and new measurements of these methods are being completed 
\cite{Abdesselam:2010pt,Altheimer:2012mn}.  To understand these methods in detail, most analyses have relied on Monte Carlo simulation 
as  the basis of study.  However, Monte Carlo simulations have limitations, and, where possible, it is vital to also compute the observables
to higher orders in QCD so as to have another handle on their behavior.

An important contribution to this effort of computing jet observables is resummation of large logarithms that arise in fixed-order
perturbation theory.  Jets are objects that are typically dominated by soft or collinear emissions and so it is necessary to resum the
logarithms that exist for an accurate prediction of an observable.  Very recently, groups have computed resummed contributions
to light jet masses at hadron colliders \cite{Li:2012bw} and N-subjettiness \cite{Thaler:2010tr,Thaler:2011gf} in color-singlet
jets at the LHC \cite{Feige:2012vc}.  Ref.~\cite{Feige:2012vc} in particular relied on the factorization of color singlet processes at hadron colliders to 
reinterpret results from $e^+e^-$ colliders.  Computing the resummed contribution to generic observables at hadron colliders
is made more difficult by the color flow throughout the collision which can destroy factorization.  To avoid discussion of these
issues, here we will only consider jet observables at $e^+e^-$ colliders.
In this paper, we will discuss the resummation of the angular correlation function introduced in \cite{Jankowiak:2011qa} using soft collinear 
effective theory (SCET) \cite{Bauer:2000ew,Bauer:2000yr,Bauer:2001ct,Bauer:2001yt}.

The angular correlation function ${\cal G}(R)$ was defined in \cite{Jankowiak:2011qa} as
\begin{equation}\label{acf_def}
{\cal G}(R)=\sum_{i\neq j}  p_{\perp i} p_{\perp j} \Delta R_{ij}^2\Theta(R-\Delta R_{ij}) \ ,
\end{equation}
for studying the substructure of jets at the LHC.  $\Delta R_{ij}$ is the boost-invariant angle between
particles $i$ and $j$, the sum runs over all constituents of a jet and $\Theta$ is the Heaviside theta function.  
The angular correlation function has distinct properties
for scaleless jets versus jets with at least one heavy mass scale.  In particular, any structure in the angular correlation function
should be distributed roughly as $R^D$, where $D$ is a constant, for a scaleless jet.
It was shown that by exploiting the different behavior of the angular distribution of hard structure in QCD jets versus jets
initiated by heavy particle decay, an efficient tagging algorithm could be defined.  

Ref.~\cite{Jankowiak:2012na} continued studying the properties of the angular correlation function, focusing on average properties
of QCD jets.  It was shown through simple calculations that, for QCD jets, the angular correlation function averaged over an ensemble of jets
should approximately scale as
\begin{equation}
\langle {\cal G}(R) \rangle \simeq R^2 \ ,
\end{equation}
where the angle brackets are defined by
\begin{equation}
\langle {\cal G}(R) \rangle = \frac{1}{N_{\text{jets}}}\sum_{i=1}^{N_{\text{jets}}} {\cal G}(R)_i \ .
\end{equation}
Deviations from $R^2$ are due to the running coupling and higher order effects.  The introduction of an ensemble averaged angular correlation
function allows for a rigorous definition of the dimension of a QCD jet which is also infrared and collinear (IRC) safe.  This dimension is defined
to be the average angular structure function $\langle \Delta {\cal G} \rangle$ and is the power to which the average angular correlation function
scales with $R$:
\begin{equation}
\langle \Delta {\cal G}(R)\rangle \equiv \frac{d\log \langle {\cal G}(R)\rangle}{d\log R} \ .
\end{equation}
For QCD jets, $\langle \Delta {\cal G} \rangle\sim 2$.  In \cite{Jankowiak:2012na}, it was also shown that the scaling of 
non-perturbative physics in $R$ is distinctly different, and this was used to determine the average energy density of the underlying event.

Here, we will continue the work of \cite{Jankowiak:2012na} and compute the average angular structure function by resummation
within the context of SCET.  Our analysis is only truly appropriate at $e^+e^-$ colliders, but we expect that the largest effect in 
going to hadron colliders is the contribution of the underlying event.  
For this calculation, we introduce generalized correlation functions ${\cal G}_\alpha(R)$ parametrized by an index $\alpha$:
\begin{equation}\label{acf_mod}
{\cal G}_\alpha(R)=\frac{1}{2E_J^2}\sum_{i\neq j} E_i E_j \sin\theta_{ij}\tan^{\alpha-1}\frac{\theta_{ij}}{2}\Theta(R-\theta_{ij}) 
\end{equation}
The form of the angular correlation function is similar to jet angularity \cite{Berger:2003iw,Almeida:2008yp}
and close in form and spirit to an event shape introduced in \cite{Banfi:2004yd}.
However, for our purposes here, we choose to index the parameter $\alpha$
such that the angular correlation function is IRC safe for all $\alpha>0$.  In the small angle limit, this reduces to Eq.~\ref{acf_def} with $\alpha=2$ (up to normalization).
The parameter $\alpha$ allows for a study of the behavior of the angular correlation function with angular scales weighted differently.
Analogously to the angular structure function, we define a generalized average angular structure function
\begin{equation}\label{asf_def}
\langle \Delta {\cal G}_\alpha \rangle \equiv \frac{d\log \langle {\cal G}_\alpha \rangle}{d\log R} \ .
\end{equation}
The calculation and interpretation of average angular structure function will be the focus of this paper.

%

In Sec.~\ref{SCET}, we discuss the factorization of jet observables in SCET and the computation of the angular correlation function
including global next-to-leading-log (NLL) contributions for jets defined by a $k_T$-type algorithm \cite{Ellis:1992qq,Catani:1993hr}.
The existence of a factorization theorem for the angular correlation function is non-trivial.  We will discuss the consistency
conditions that the angular correlation function satisfies for factorization.  We will also briefly discuss how the results obtained here can be used in a 
calculation of the angular correlation function at the LHC.  
In Sec.~\ref{FOsec} we compare the SCET calculation to a next-to-leading-order (NLO) calculation of the angular correlation function.  Resummation
and fixed-order corrections affect different parts of distributions and so the differences between the resummed calculation and the fixed-order result
give some sense as to the importance of these effects.
This analysis leads to Sec.~\ref{MC}, were we present a comparison between the SCET calculation 
and the output of parton shower Monte Carlo.  We observe significant differences between SCET and Monte Carlo, but higher fixed order effects are substantial.  
We discuss some of the uncertainties in the parton shower studying the effect of the evolution variable on the value of the angular structure function.
  Finally, we present our conclusions in Sec.~\ref{conc}.

\section{\label{SCET}SCET Calculation}

SCET  is an effective theory of QCD in which all modes of QCD are integrated out
except those corresponding to soft or collinear modes.  Collinear and soft modes are
defined by their scaling with power counting parameter $\lambda$:
\begin{eqnarray*}
\text{collinear}&\sim& (\lambda^2,1,\lambda)  \ ,\\
\text{soft}&\sim& (\lambda^2,\lambda^2,\lambda^2) \ ,
\end{eqnarray*}
which is the scaling of the $+$, $-$ and transverse components of the momenta, respectively.  $\lambda$ is a parameter
that is defined for a particular process or observable; for example, for computing the distribution of jet masses, 
$\lambda\sim m_J / p_{\perp J}\ll1$.  The fact that $\lambda \ll 1$ allows for a systematic expansion in powers
of $\lambda$.  Higher order terms in $\lambda$ are power suppressed (much like the subleading terms in the twist expansion).

For an event shape observable ${\cal O}$ that factorizes, the cross section can be written in the schematic form:
\begin{equation}
\frac{d\sigma}{d{\cal O}} = H(\mu) \left[ \prod_{n_i} J_{n_i}({\cal O};\mu) \right]\otimes S({\cal O};\mu) \ ,
\end{equation}
where $H(\mu)$ is the hard function, which matches the full QCD result at a scale $\mu$, $J(\mu;n_i,{\cal O})$ is the jet function
for the contribution to the observable ${\cal O}$ from $n_i$-collinear modes and $S(\mu;{\cal O})$ 
is the soft function for the contribution to the observable ${\cal O}$ from the soft modes.  $\otimes$ represents a convolution
between the jet and soft functions.
  All functions depend on the factorization scale $\mu$.  

Factorization of jet observables in SCET was first exhibited in \cite{Ellis:2009wj,Ellis:2010rwa}.
 Ref.~\cite{Ellis:2010rwa} computed
individual jet angularities to NLL in $e^+e^-$ collisions.  It was shown that factorization of the cross section for jet observables 
in $e^+e^-\to N$ jets has the form
\begin{eqnarray}
\frac{d\sigma}{d{\cal O}_1\cdots d{\cal O}_M}= H(n_1,\ldots,n_N;\mu)\left[ \prod_{i=1}^M J_{n_i}({\cal O}_i;\mu)\right]
\otimes S_{n_1\cdots n_N}({\cal O}_1,\ldots,{\cal O}_M;\mu) \prod_{j=M+1}^N J(\mu)\ , \ \ 
\end{eqnarray}
where $M\leq N$ of the jet observables ${\cal O}_i$ have been measured.  Jet directions are denoted by $n_i$ and
$J({\cal O}_i;\mu)$ is the jet function for  a jet in which the observable ${\cal O}_i$ has been measured and $J(\mu)$
is a jet function for a jet which has not been measured.  We will refer to these as the measured and unmeasured jet functions,
respectively.  A similar nomenclature will be used for the soft functions.
Jet algorithm dependence and jet energies have been suppressed.  An important point from \cite{Ellis:2010rwa} is that
factorization requires that the jets be well-separated; namely, that
\begin{equation}\label{tsup}
t_{ij} = \frac{\tan\frac{\psi_{ij}}{2}}{\tan\frac{R_0}{2}} \gg 1 \ ,
\end{equation}
where $\psi_{ij}$ the is angle between any pair of jets $i,j$ and $R_0$ is the jet algorithm radius.  We will assume that this condition 
is met in the following and leave any discussion of subtleties to \cite{Ellis:2010rwa}.
A non-trivial requirement of the factorization is the independence of the cross section on the factorization scale $\mu$.  This requirement
leads to a constraint that the sum of the anomalous dimensions of the hard, jet and soft functions is zero.
We will show that this holds for the angular correlation function.

We will use the results of the factorization theorem proven in \cite{Ellis:2009wj,Ellis:2010rwa} 
 to compute 
the distribution of the angular correlation function from Eq.~\ref{acf_mod}.  In particular, we are interested in the ensemble average
of the angular structure function as defined in Eq.~\ref{asf_def}.  Note that this observable is independent of any normalization factor
of the angular correlation function; thus, with the goal of computing the average angular structure function, it is consistent to ignore
factors that are independent of ${\cal G}_\alpha$ and the angular resolution parameter $R$.  Thus, for the purposes of this paper, we
can ignore the overall factors in the factorized form of the cross section of the hard function and the unmeasured jet functions.  In this
case, the factorized form of the cross section becomes
\begin{eqnarray}
\frac{d\sigma}{d{\cal G}_{\alpha1}\cdots d{\cal G}_{\alpha M}}= C(\mu)\left[ \prod_{i=1}^M J_{n_i}({\cal G}_{\alpha i};\mu)\right]
\otimes S_{n_1\cdots n_N}({\cal G}_{\alpha1},\ldots,{\cal G}_{\alpha M};\mu) \ ,
\end{eqnarray}
where $C(\mu)$ is independent of ${\cal G}_\alpha$ and the resolution parameter $R$.

In this section, we present a calculation of the jet and soft functions for the angular correlation function for jets defined by 
a $k_T$ algorithm.  We first argue that the angular correlation function is computable in SCET and relate its form at NLO to the
form of jet angularity at NLO.  This comparison will allow us to relate the calculation of the angular correlation function to the work 
in \cite{Ellis:2010rwa}.  We then present a calculation of the measured jet and soft functions of the angular correlation 
function.  From these results, we can determine the anomalous dimensions of the jet and soft functions and will show the
consistency of the factorization relies on a non-trivial cancellation of dependence on the angular resolution $R$ between the 
jet and soft functions.  We can then resum up to the next-to-leading logs of the jet and soft functions by the renormalization group.
Note that we do not attempt to resum non-global logs \cite{Dasgupta:2001sh} 
that arise due to the non-trivial phase space
constraints of the jet algorithm or the angular correlation function.
  From the resummed expression of the angular correlation function,
we find the ensemble average and compute the average angular structure function numerically.

It should be stressed that non-global logarithms are ignored in this study.  The angular correlation function for a jet requires several phase space
constraints; the jet algorithm, soft jet vetoes, the resolution parameter $R$, {\it etc}.  These provide numerous sources for non-global logarithms 
which cannot be resummed analytically.  The study of non-global logarithms in QCD cross sections is a subtle and evolving story.  For recent work in this direction, 
especially in the context of non-global logarithms from jet clustering
see, for example, \cite{Banfi:2005gj,Delenda:2006nf,Hornig:2011tg,KhelifaKerfa:2011zu,Kelley:2012kj,Kelley:2012zs}.
It is outside the scope of this paper to discuss non-global logarithms further.

\subsection{\label{jetfacto}Factorization of the Angular Correlation Function}

Factorization of jet observables requires that soft modes only resolve 
the entire jet and not individual collinear modes contributing to the jet.  Angularity $\tau_a$ is a one-parameter family
of observables defined as \cite{Berger:2003iw,Almeida:2008yp}
\begin{equation}
\tau_a = \frac{1}{2E_J}\sum_{i\in J} e^{-\eta_i(1-a)} p_{\perp i} \ , 
\end{equation}
where $J$ is the jet, $p_{\perp i}$ is the momentum of particle $i$ transverse to the jet axis and $\eta_i$ is the rapidity of particle $i$
with respect to the jet axis:
\begin{equation}
\eta_i = -\log \tan \frac{\theta_i}{2} \ .
\end{equation}
Angularity is IRC safe for $a<2$.  The separation of soft and collinear modes in angularity is simple to show.  To leading power in $\lambda$,
\begin{eqnarray}
\tau_a &=& \frac{1}{2E_J}\sum_{C\in J} e^{-\eta_C(1-a)} p_{\perp C} +  \frac{1}{2E_J}\sum_{S\in J} e^{-\eta_S(1-a)} p_{\perp S} \nonumber \\
&=& \tau_a^C + \tau_a^S \ ,
\end{eqnarray}
where $C$ and $S$ represent the collinear and soft modes, respectively.  Note that the soft modes do not affect the location of the jet center 
to leading power in $\lambda$.
  Factorization of angularities exists only for $a<1$ due to
the presence of logarithms of rapidity; however,
recently it was shown that these logarithms can be controlled \cite{Chiu:2011qc,Chiu:2012ir}.
We will show that angularity and the angular correlation function have similarities which 
will allow us to use many of the results from \cite{Ellis:2010rwa} here.




To justify the use of SCET for computing the angular correlation function, we must first show that the angular correlation function
does not mix soft and collinear modes.  This argument was presented in \cite{Jankowiak:2012na} (based on arguments from \cite{Walsh:2011fz}), 
but we present it here for completeness.
In terms of soft and collinear modes, the angular correlation function can be expressed as
\begin{eqnarray}
{\cal G}_\alpha(R)&=&\frac{1}{2E_J^2}\sum_{i\neq j} E_i E_j \sin\theta_{ij}\tan^{\alpha-1}\frac{\theta_{ij}}{2}\Theta(R-\theta_{ij}) \nonumber\\
&=&\frac{1}{2E_J^2}\sum_{i, j\in C} E_i E_j \sin\theta_{ij}\tan^{\alpha-1}\frac{\theta_{ij}}{2}\Theta(R-\theta_{ij}) \nonumber\\
&&+ \ \frac{1}{2E_J^2}\sum_{i, j\in S} E_i E_j \sin\theta_{ij}\tan^{\alpha-1}\frac{\theta_{ij}}{2}\Theta(R-\theta_{ij}) \nonumber\\
&& + \ \frac{1}{2E_J^2}\sum_{C, S} E_C E_S \sin\theta_{CS}\tan^{\alpha-1}\frac{\theta_{CS}}{2}\Theta(R-\theta_{CS}) \ . 
\end{eqnarray}
Note that, to NLO, there is no soft-soft correlation contribution to the angular correlation
function because such a term would require the radiation of two soft gluons which first occurs at NNLO.  To accuracy of the leading power in $\lambda$, we can
exchange the collinear modes with the jet itself in the collinear-soft term.  Explicitly,
\begin{equation}
\theta_{CS} = \theta_{JS} + {\cal O}(\lambda)\ , 
\end{equation}
as the angle of the soft modes with respect to the jet center scales as $\theta_{JS}\sim 1$.
Appropriate for NLO or NLL, the angular
correlation function can be written as
\begin{eqnarray}
{\cal G}_\alpha(R)
&=&\frac{1}{2E_J^2}\sum_{i, j\in C} E_i E_j \sin\theta_{ij}\tan^{\alpha-1}\frac{\theta_{ij}}{2}\Theta(R-\theta_{ij}) \nonumber\\
&& + \ \frac{1}{2E_J}\sum_{S}  E_S \sin\theta_{JS}\tan^{\alpha-1}\frac{\theta_{JS}}{2}\Theta(R-\theta_{JS}) \ .
\end{eqnarray}
Thus, the collinear and soft modes are decoupled to leading power and so the angular correlation function is factorizable, and hence
computable, in SCET.

To NLO, a jet is composed of at most two particles, so the form of many observables simplifies substantially at this order.  
The form of the angular correlation function from Eq.~\ref{acf_mod} was chosen so as to be similar in form to angularity.  The contribution
to the angularity and the angular correlation function from collinear modes is distinct. The measured jet functions will
need to be recomputed for the angular correlation function.  However, the contributions to the angularity and the angular
correlation function from soft modes are simply related:
\begin{equation}
{\cal G}_\alpha^{S}(R)= \frac{E_S}{2 E_J} \sin \theta_{JS} \tan^{\alpha-1}\frac{\theta_{JS}}{2}\Theta(R-\theta_{JS})=\tau_{2-\alpha}^{S}\Theta(R-\theta_{JS}) \ .
\end{equation}
This observation will allow us to recycle the soft function calculation for angularity for the angular correlation function.

An important point to note here is that the scaling of the angle between collinear modes $i$ and $j$ goes like $\theta_{ij}\sim \lambda$.  Thus,
to leading power, the angular correlation function for the collinear-collinear contribution can be written as
\begin{eqnarray}
{\cal G}_\alpha^{CC} &=& \frac{1}{2E_J^2}\sum_{i, j\in C} E_i E_j \sin\theta_{ij}\tan^{\alpha-1}\frac{\theta_{ij}}{2}\Theta(R-\theta_{ij}) \nonumber\\
&=&\frac{1}{E_J^2}\sum_{i, j\in C} E_i E_j \tan^{\alpha}\frac{\theta_{ij}}{2}\Theta(R-\theta_{ij}) \ . 
\end{eqnarray}
We will use this form of the collinear-collinear contribution to the angular correlation function for computing the measured jet functions.

Throughout this paper, we will only consider jets with a single collinear sector.  Small values of the angular correlation function are
not enough to guarantee that the jet has only a single collinear sector, however, we believe that contributions from multiple collinear sectors is subdominant.
Our reasoning is as follows.  First, at large values of $R\sim R_0$, the angular correlation function is essentially
an angular-weighted jet mass measure.  In this case, additional collinear sectors would be correlated increasing the value of the angular
correlation function substantially.  At small $R$, there
are two options: either collinear sectors are still correlated or they are uncorrelated.  If the collinear sectors are still correlated at small $R$
(they are within an angle $R$ of one another), then logarithms of this angle will appear.  However, the logarithms of the angle between the separate
collinear modes should be subdominant to the logarithms of the resolution parameter of the angular correlation function, $R$.  
We do not attempt to resum the latter logarithms in this paper.
If the collinear sectors are uncorrelated and separated by an angle larger than $R$, then the jet effectively breaks up into several smaller jets each with 
similar scaling properties as $R\to 0$.  In sum, we expect the effect of additional collinear sectors to be significantly subdominant so as to be
consistently ignored in this study.  We believe the absence of fixed-order terms is a more important omission.

\subsection{\label{measjet}Measured Jet Functions}

The leading power contribution to the measured jet functions at NLO comes from two collinear particles which are clustered in the jet
and can be computed from cutting one-loop SCET diagrams.  The phase space integrals can be extended over the entire
range of  momentum for the collinear particles in the jet as long as the contribution from the zero momentum bin
is subtracted \cite{Manohar:2006nz}.  In particular, we consider a jet with light cone momentum $l=(l^+,\omega,0)$ which splits
to two collinear particles with light cone momenta $q=(q^+,q^-,{\bf q}_\perp)$ and $l-q=(l^+-q^+,\omega-q^-,-{\bf q}_\perp)$.  
The zero-bin subtraction
term can be determined from the measured jet function by taking the scaling $q\sim \lambda^2$.  We will refer to contribution to the jet function
that does not include the zero-bin subtraction as the na\"ive contribution.

To compute the measured jet function, we will need to enforce phase space cuts from the jet algorithm and the observable.  We will
compute the jet function for a $k_T$-type jet algorithm  as defined by a jet radius $R_0$.  At NLO, all $k_T$ algorithms are the
same and two particles are clustered in the jet if their angular separation is less than $R_0$.  This leads to the phase space constraint
\begin{equation}
\Theta_{k_T}=\Theta\left( \cos R_0 - \frac{{\bf q}\cdot ({\bf l}-{\bf q})}{|{\bf q}|\sqrt{({\bf l}-{\bf q})^2}}  \right) = \Theta\left( \tan^2\frac{R_0}{2} - \frac{q^+ \omega^2}{q^-(\omega-q^-)^2}   \right) \ ,
\end{equation}
where on the right, the leading scaling behavior was kept.  The jet algorithm constraint for the zero-bin subtraction term is then
\begin{equation}
\Theta_{k_T}^{(0)} = \Theta\left( \tan^2\frac{R_0}{2} - \frac{q^+ }{q^-}   \right) \ . 
\end{equation}

The phase space constraints for the angular correlation function are more subtle.  The $\delta$-function which constrains a jet to have angular correlation function ${\cal G}_\alpha$, $\delta_R = \delta({\cal G}_\alpha - \hat{\cal G}_\alpha)$, is
\begin{equation}
\delta_R = \delta\left({\cal G}_\alpha-\omega^{\alpha-2}(\omega-q^-)^{1-\alpha}(q^-)^{1-\alpha/2}(q^+)^{\alpha/2}\Theta\left(\tan^2  \frac{R}{2}  - \frac{q^+ \omega^2}{q^- (\omega -q^-)^2}\right)\right) \ ,
\end{equation} 
where $R$ is the resolution parameter of the angular correlation function.  For a $k_T$-type jet at NLO, the angular correlation
function vanishes if $R>R_0$; thus, we will assume that $R<R_0$ in the following.
This $\delta$-function can be decomposed depending on the value of $\Theta$-function as
\begin{eqnarray}
\delta_R &= &  \ \delta\left({\cal G}_\alpha-\omega^{\alpha-2}(\omega-q^-)^{1-\alpha}(q^-)^{1-\alpha/2}(q^+)^{\alpha/2}\Theta\left(\tan^2  \frac{R}{2}  - \frac{q^+ \omega^2}{q^- (\omega -q^-)^2}\right)\right) \nonumber \\
 &= & \  \delta\left({\cal G}_\alpha-\omega^{\alpha-2}(\omega-q^-)^{1-\alpha}(q^-)^{1-\alpha/2}(q^+)^{\alpha/2}\right)\Theta\left(\tan^2 \frac{R}{2}  - \frac{q^+ \omega^2}{q^- (\omega -q^-)^2}\right) \nonumber \\
 && + \ \delta\left({\cal G}_\alpha\right)\Theta\left(\frac{q^+ \omega^2}{q^- (\omega -q^-)^2}-\tan^2  \frac{R}{2}  \right) \ .
\end{eqnarray}
The $\delta$-function for the zero-bin subtraction term is found by taking $q\sim \lambda^2$:
\begin{equation}
\delta_R^{(0)} = \delta\left({\cal G}_\alpha-\omega^{-1}(q^-)^{1-\alpha/2}(q^+)^{\alpha/2}\right)\Theta\left(\tan^2 \frac{R}{2}  - \frac{q^+ }{q^-}\right)
 +  \delta\left({\cal G}_\alpha\right)\Theta\left(\frac{q^+}{q^- }-\tan^2 \frac{R}{2}  \right) \ .
\end{equation}

\subsubsection{\label{quarkmeasfunc}Measured Quark Jet Function}

\begin{figure}
 \includegraphics[width=16cm]{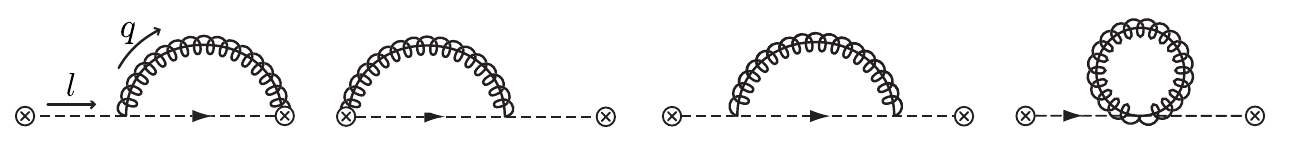}%
 \caption{\label{fig:quark_diag}SCET Feynman diagrams contributing to the quark jet function.}
 \end{figure}

The na\"{i}ve contribution to the measured quark jet function can be computed in dimensional regularization from the diagrams
shown in Fig.~\ref{fig:quark_diag}:
\begin{eqnarray}
\tilde{J}_{\omega}^{q}({\cal G}_\alpha) &=& \  g^2\mu^{2\epsilon}C_F \int \frac{dl^+}{2\pi}\frac{1}{(l^+)^2}
\int \frac{d^d q}{(2\pi)^d}\left( 4\frac{l^+}{q^-} +(d-2) \frac{l^+-q^+}{\omega-q^-}  \right) \nonumber\\
&&\times 2\pi \delta\left(q^+q^--q_\perp^2\right) \Theta(q^-)\Theta(q^+)2\pi \delta\left( l^+ - q^+ - \frac{q_\perp^2}{\omega-q^-}\right)\nonumber\\
&&\times \Theta(\omega-q^-)\Theta(l^+-q^+) \Theta\left(  \tan^2\frac{R_0}{2} - \frac{q^+\omega^2}{q^-(\omega-q^-)^2} \right)\nonumber\\
&&\times\left[ \delta\left({\cal G}_\alpha-\omega^{\alpha-2}(\omega-q^-)^{1-\alpha}(q^-)^{1-\alpha/2}(q^+)^{\alpha/2}\right)\Theta\left(\tan^2  \frac{R}{2} - \frac{q^+ \omega^2}{q^- (\omega -q^-)^2}\right)\right. \nonumber \\
  && + \left. \delta\left({\cal G}_\alpha\right)\Theta\left(\frac{q^+ \omega^2}{q^- (\omega -q^-)^2}-\tan^2 \frac{R}{2} \right) \right] \ .
\end{eqnarray}
We take $d=4-2\epsilon$.
The coefficient to the $\delta({\cal G}_\alpha)$ term can be found by integrating over ${\cal G}_\alpha$.  The terms that remain
are $+$-distributions, which integrate to zero.  The zero-bin subtraction term follows from taking the scaling 
limit $q\sim\lambda^2$ of the na\"{i}ve jet function above:
\begin{eqnarray}
J_{\omega}^{q(0)}({\cal G}_\alpha) &=& \  4g^2\mu^{2\epsilon}C_F \int \frac{dl^+}{2\pi}\frac{1}{l^+}
\int \frac{d^d q}{(2\pi)^d}\frac{1}{q^-}   2\pi \delta\left(q^+q^--q_\perp^2\right) \Theta(q^-)\Theta(q^+)\nonumber \\
&&\times2\pi \delta\left( l^+ - q^+\right) \Theta(l^+-q^+) \Theta\left(  \tan^2\frac{R_0}{2} - \frac{q^+}{q^-} \right)\nonumber \\
&&\times\left[ \delta\left({\cal G}_\alpha-\omega^{-1}(q^-)^{1-\alpha/2}(q^+)^{\alpha/2}\right)\Theta\left(\tan^2  \frac{R}{2} - \frac{q^+ }{q^- }\right)\right.\nonumber \\
&&+\left.  \delta\left({\cal G}_\alpha\right)\Theta\left(\frac{q^+}{q^- }-\tan^2 \frac{R}{2} \right) \right] \ .
\end{eqnarray}
The term proportional to $\delta({\cal G}_\alpha)$ is scaleless and integrates to zero in pure dimensional regulation.

Employing a $\overline{\text{MS}}$ scheme, we find the measured quark jet function for $k_T$-type jet algorithms
of ${\cal G}_\alpha$ to be
\begin{eqnarray}\label{Qjetfunc}
J_{\omega}^{q}({\cal G}_\alpha)=\tilde{J}_{\omega}^{q}({\cal G}_\alpha)-J_{\omega}^{q(0)}({\cal G}_\alpha) &=& \frac{\alpha_s C_F}{2\pi}\left[ \left( \frac{\alpha}{\alpha-1}\frac{1}{\epsilon^2} +\frac{3}{2}\frac{1}{\epsilon}+\frac{\alpha}{\alpha-1}\frac{\log\frac{\mu^2}{\omega^2}}{\epsilon}+\frac{1}{\epsilon}\log\frac{\tan^2\frac{R}{2}}{\tan^2\frac{R_0}{2}} \right)\delta({\cal G}_\alpha)\right.\nonumber \\
&&\left.- \ \frac{2}{\alpha-1}\frac{1}{\epsilon}\left( \frac{\Theta({\cal G}_\alpha)}{{\cal G}_\alpha} \right)_+\right] + J_{\omega}^{q}({\cal G}_\alpha,\epsilon^0) \ ,
\end{eqnarray}
where $J_{\omega}^{q}({\cal G}_\alpha,\epsilon^0)$ consists of terms that are finite as $\epsilon\to 0$. 
  These terms are presented in
Appendix~\ref{jet_func_app}. The definition of the 
$+$-distribution is also given in Appendix~\ref{jet_func_app}.
  Note that the $1/\epsilon$ terms for the angular correlation function are the same as those for angularity
from \cite{Ellis:2010rwa} with $\alpha\to 2-a$ plus an additional term of the logarithm of the ratio of scales; the resolution scale $R$ and
the jet radius $R_0$.  This term contributes to the anomalous dimension of the jet function. In principle, these logarithms
 could be attempted
to be resummed.  However, note that the resolution scale $R$ can never practically be parametrically smaller than the jet radius $R_0$,
so these logarithms never become large.  Thus, we will not worry about resumming these logarithms.

\subsubsection{\label{gluonmeasfunc}Measured Gluon Jet Function}

 \begin{figure}
 \includegraphics[width=16cm]{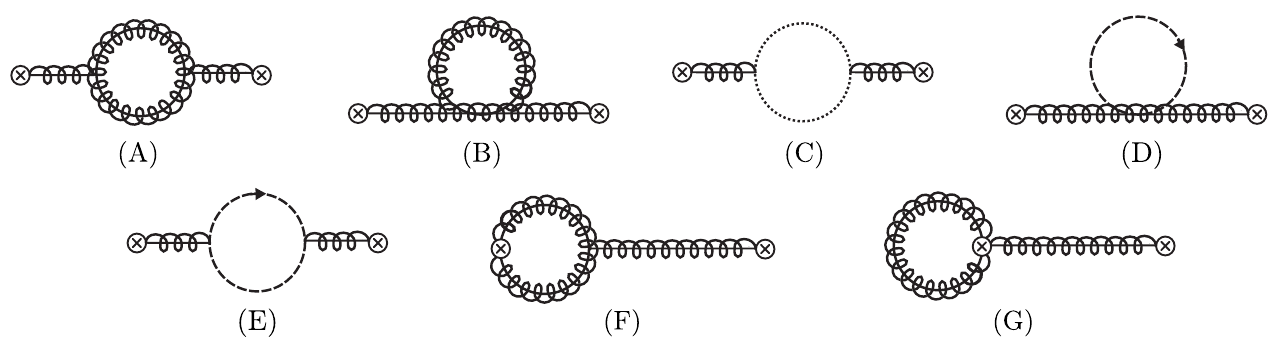}%
 \caption{\label{fig:gluon_diag}SCET Feynman diagrams contributing to the gluon jet function.  Diagrams (F) and (G) have
 mirrored counterparts which are not shown.}
 \end{figure}

The na\"{i}ve contribution to the measured gluon jet function can be computed from the diagrams
shown in Fig.~\ref{fig:gluon_diag}:
\begin{eqnarray}
\tilde{J}_{\omega}^{g}({\cal G}_\alpha) &=& \  2g^2\mu^{2\epsilon} \int \frac{dl^+}{2\pi}\frac{1}{l^+}
\int \frac{d^d q}{(2\pi)^d}  \frac{1}{\omega-q^-}  2\pi \delta\left(q^+q^--q_\perp^2\right) 2\pi \delta\left( l^+ - q^+ - \frac{q_\perp^2}{\omega-q^-}\right) \nonumber\\
&&\times \left\{ n_F T_R \left(1-\frac{2}{1-\epsilon}\frac{q^+q^-}{\omega l^+}  \right) - C_A \left( 2-\frac{\omega}{q^-}-\frac{\omega}{\omega-q^-} - \frac{q^+q^-}{\omega l^+} \right)  \right\}\nonumber\\
&&\times \Theta(q^-)\Theta(q^+)\Theta(\omega-q^-)\Theta(l^+-q^+) \Theta\left(  \tan^2\frac{R_0}{2} - \frac{q^+\omega^2}{q^-(\omega-q^-)^2} \right)\nonumber\\
&&\times\left[ \delta\left({\cal G}_\alpha-\omega^{\alpha-2}(\omega-q^-)^{1-\alpha}(q^-)^{1-\alpha/2}(q^+)^{\alpha/2}\right)\Theta\left(\tan^2  \frac{R}{2} - \frac{q^+ \omega^2}{q^- (\omega -q^-)^2}\right)\right. \nonumber \\
  && + \left. \delta\left({\cal G}_\alpha\right)\Theta\left(\frac{q^+ \omega^2}{q^- (\omega -q^-)^2}-\tan^2 \frac{R}{2} \right) \right] \ .
\end{eqnarray}
The coefficient of the $\delta({\cal G}_\alpha)$ term can be found by integrating over ${\cal G}_\alpha$.  The terms that remain
are $+$-distributions, which integrate to zero.  The zero-bin subtraction term follows from taking the scaling 
limit $l-q\sim q\sim\lambda^2$ of the na\"{i}ve jet function above:
\begin{eqnarray}
J_{\omega}^{g(0)}({\cal G}_\alpha) &=& \  4g^2\mu^{2\epsilon}C_A \int \frac{dl^+}{2\pi}\frac{1}{l^+}
\int \frac{d^d q}{(2\pi)^d}\frac{1}{q^-}   2\pi \delta\left(q^+q^--q_\perp^2\right) \Theta(q^-)\Theta(q^+)\nonumber \\
&&\times2\pi \delta\left( l^+ - q^+\right) \Theta(l^+-q^+) \Theta\left(  \tan^2\frac{R_0}{2} - \frac{q^+}{q^-} \right)\nonumber \\
&&\times\left[ \delta\left({\cal G}_\alpha-\omega^{-1}(q^-)^{1-\alpha/2}(q^+)^{\alpha/2}\right)\Theta\left(\tan^2  \frac{R}{2} - \frac{q^+ }{q^- }\right)\right.\nonumber \\
&&+\left.  \delta\left({\cal G}_\alpha\right)\Theta\left(\frac{q^+}{q^- }-\tan^2 \frac{R}{2} \right) \right] \ .
\end{eqnarray}
The term proportional to $\delta({\cal G}_\alpha)$ integrates to zero in pure dimensional regulation.  This zero-bin subtraction term is exactly the same up to color 
factors as the quark jet function zero-bin subtraction.  

Employing a $\overline{\text{MS}}$ scheme, we find the measured gluon jet function for the $k_T$-type jet algorithms
of ${\cal G}_\alpha$ to be
\begin{eqnarray}\label{Gjetfunc}
J_{\omega}^{g}({\cal G}_\alpha)=\tilde{J}_{\omega}^{g}({\cal G}_\alpha)-J_{\omega}^{g(0)}({\cal G}_\alpha) &=& \frac{\alpha_s}{2\pi}\left[ 
\left(C_A\frac{\alpha}{\alpha-1}\frac{1}{\epsilon^2}+\frac{\beta_0}{2\epsilon}+C_A\frac{\alpha}{\alpha-1}\frac{\log \frac{\mu^2}{\omega^2}}{\epsilon} \right.\right.\nonumber\\
&&\left.\left.+ \ \frac{C_A}{\epsilon}\log\frac{\tan^2\frac{R}{2}}{\tan^2\frac{R_0}{2}} \right)\delta({\cal G}_\alpha)
 -\frac{2C_A}{\epsilon(\alpha-1)}\left( \frac{\Theta({\cal G}_\alpha)}{{\cal G}_\alpha} \right)_+
\right] \nonumber\\
&&+ \  J_{\omega}^{g}({\cal G}_\alpha,\epsilon^0) \ ,
\end{eqnarray}
where $J_{\omega}^{g}({\cal G}_\alpha,\epsilon^0)$ consists of terms that are finite as $\epsilon\to 0$.  These terms are presented in 
Appendix~\ref{jet_func_app}.  $\beta_0$ is the coefficient of the one-loop $\beta$-function:
\begin{equation}\label{beta_one}
\beta_0 = \frac{11}{3}C_A - \frac{2}{3}N_{F} \ ,
\end{equation}
with $T_R = \frac{1}{2}$.
As with the quark jet function, the 
$1/\epsilon$ terms are the same as those for angularity
from \cite{Ellis:2010rwa} with $\alpha\to 2-a$ plus an additional term of the logarithm of the ratio of the resolution parameter $R$ to the 
jet radius $R_0$.

\subsection{\label{meassoft}Measured Soft Function}

As shown above, there is a simple relationship between the form of angularity for soft modes and the angular correlation for soft modes.  
This relationship will allow us to use the results from \cite{Ellis:2010rwa} in computing the measured soft function for
the angular correlation function.  First, we consider the phase space constraints from the jet algorithm and the angular correlation
function.  For the $k_T$ jet algorithm, soft radiation must be within the jet radius $R_0$ of the jet axis to be included:
\begin{equation}
\Theta_{k_T}=\Theta\left(\tan^2\frac{R_0}{2}-\frac{k^+}{k^-}\right) \ .
\end{equation}
The $\delta$-function that constrains the soft modes to contribute an amount ${\cal G}_a$ to the angular correlation function
is 
\begin{eqnarray}
\delta_R &=& \delta\left({\cal G}_\alpha-\omega^{-1}(k^-)^{1-\alpha/2}(k^+)^{\alpha/2}\Theta\left(\tan^2 \frac{R}{2}  - \frac{k^+ }{k^-}\right)\right)  \nonumber \\ 
&=& \delta\left({\cal G}_\alpha-\omega^{-1}(k^-)^{1-\alpha/2}(k^+)^{\alpha/2}\right)\Theta\left(\tan^2 \frac{R}{2}  - \frac{k^+ }{k^-}\right) + \delta\left({\cal G}_\alpha\right)\Theta\left(\frac{k^+}{k^- }-\tan^2  \frac{R}{2}  \right) \ .
\end{eqnarray}

The measured soft function of a gluon emitted from lines $i$ and $j$ into a jet is
\begin{eqnarray}
S_{ij}^{\text{meas}}({\cal G}_\alpha) &=& -g^2 \mu^{2\epsilon} {\bf T}_i\cdot {\bf T}_j \int \frac{d^dk}{(2\pi)^d}\frac{n_i\cdot n_j}{(n_i\cdot k)(n_j\cdot k)}
2\pi\delta(k^2)\Theta(k^0)\Theta_{k_T}\delta_R \nonumber \\
&=& -g^2 \mu^{2\epsilon} {\bf T}_i\cdot {\bf T}_j \int \frac{d^dk}{(2\pi)^d}\frac{n_i\cdot n_j}{(n_i\cdot k)(n_j\cdot k)}
2\pi\delta(k^2)\Theta(k^0)\Theta\left( \tan^2\frac{R_0}{2}-\frac{k^+}{k^-}  \right)\nonumber\\
&&\times \left[\delta\left({\cal G}_\alpha-\omega^{-1}(k^-)^{1-\alpha/2}(k^+)^{\alpha/2}\right)\Theta\left(\tan^2 \frac{R}{2}  - \frac{k^+ }{k^-}\right)\right.  \nonumber\\
&&\left.+ \  \delta\left({\cal G}_\alpha\right)\Theta\left(\frac{k^+}{k^- }-\tan^2  \frac{R}{2}  \right) \right]  \ .
\end{eqnarray}
Note that the integral proportional to $\delta({\cal G}_\alpha)$ is scaleless and so vanishes in pure dimensional regularization.  Also, the integral
is only non-zero if $R<R_0$ and so the $\Theta$-function from the jet algorithm is redundant.
  Thus, we can write the soft function as
\begin{eqnarray}
S_{ij}^{\text{meas}}({\cal G}_\alpha) 
&=& -g^2 \mu^{2\epsilon} {\bf T}_i\cdot {\bf T}_j \int \frac{d^dk}{(2\pi)^d}\frac{n_i\cdot n_j}{(n_i\cdot k)(n_j\cdot k)}
2\pi\delta(k^2)\Theta(k^0)\Theta\left( \tan^2\frac{R}{2}-\frac{k^+}{k^-}  \right)\nonumber\\
&&\times \  \delta\left({\cal G}_\alpha-\omega^{-1}(k^-)^{1-\alpha/2}(k^+)^{\alpha/2}\right)  \ .
\end{eqnarray}
This is the same form of the measured soft function as for angularity with a jet radius equal to $R$ which was
computed in \cite{Ellis:2010rwa}.  Up to terms that are suppressed by $1/t^2$ from Eq.~\ref{tsup}, the measured soft function
for jet $i$ is
\begin{eqnarray}\label{soft_func}
S^{\text{meas}}({\cal G}_{i\alpha})&=&-\frac{\alpha_s}{2\pi}{\bf T}_i^2\frac{1}{\alpha-1}\left\{ \left[ \frac{1}{\epsilon^2}+\frac{1}{\epsilon}\log\frac{\mu^2\tan^{2(\alpha-1)}\frac{R}{2}}{\omega^2}-\frac{\pi^2}{12}+\frac{1}{2}\log^2\frac{\mu^2\tan^{2(\alpha-1)}\frac{R}{2}}{\omega^2}  \right] \delta({\cal G}_{i\alpha})\right.\nonumber\\
&&\left. -2\left[ \left( \frac{1}{\epsilon} + \log\frac{\mu^2\tan^{2(\alpha-1)}\frac{R}{2}}{{\cal G}_{i\alpha}^2\omega^2}  \right)\frac{\Theta({\cal G}_{i\alpha})}{{\cal G}_{i\alpha}}  \right]_+ \right\} \ ,
\end{eqnarray}
where ${\bf T}_i^2$ is the square of the color in the jet.

\subsection{\label{anomdim}Anomalous Dimensions and Consistency Conditions}

A non-trivial requirement of the factorization is that the physical cross section should be independent of the factorization scale $\mu$.  A
consequence of this is that the anomalous dimensions of the hard, jet and soft functions must sum to 0.  The requirement is
\begin{equation}\label{anom_sum}
0=\left( \gamma_H(\mu) + \gamma_S^{\text{unmeas}}(\mu)+\sum_{i\notin \text{meas}}\gamma_{J_i}(\mu) \right)\delta({\cal G}_\alpha)+\sum_{i\in\text{meas}}\left(  \gamma_{J_i}({\cal G}_\alpha^i;\mu)+\gamma_S^{\text{meas}}({\cal G}_\alpha^i;\mu)   \right) \ ,
\end{equation}
where $\gamma_H$, $\gamma_S$ and $\gamma_J$ are the anomalous dimensions of the hard, soft and jet functions.  
The $\mu$ dependence must be summed over the measured and unmeasured jet and soft functions.
The sum of the hard, unmeasured soft and unmeasured jet anomalous dimensions to NLO is
\begin{equation}\label{unmeas_anom}
 \gamma_H(\mu) + \gamma_S^{\text{unmeas}}(\mu)+\sum_{i\notin \text{meas}}\gamma_{J_i}(\mu) = -\frac{\alpha_s}{\pi}\sum_{i\in \text{meas}} {\bf T}_i^2 \log\frac{\mu^2}{\omega_i^2\tan^2\frac{R_0}{2}}-\sum_{i\in\text{meas}}\gamma_i \ ,
\end{equation}
where $\gamma_i$ depends on the flavor of the jet:
\begin{equation}\label{gamma_flav}
\gamma_q=\frac{3\alpha_s}{2\pi}C_F \ , \qquad \gamma_g=\frac{\alpha_s}{\pi}\frac{11C_A-2N_F}{6} =\frac{\alpha_s}{2\pi}\beta_0\ ,
\end{equation}
for quark and gluon jets respectively.  We will show that the measured jet and soft function anomalous dimensions for the 
angular correlation function are exactly what is required to satisfy Eq.~\ref{anom_sum}.

The anomalous dimensions of the measured jet or soft functions are given by the coefficient of the $1/\epsilon$ terms from
Eqs.~\ref{Qjetfunc}, \ref{Gjetfunc} and \ref{soft_func}.  The anomalous dimensions of the quark and gluon jet functions can be written
collectively as
\begin{equation}\label{jetmeasad}
\gamma_{J_i}({\cal G}_\alpha^i) =  \left[\frac{\alpha_s}{\pi}{\bf T}_i^2\left( \frac{\alpha}{\alpha-1}\log\frac{\mu^2}{\omega_i^2} +\log\frac{\tan^2\frac{R}{2}}{\tan^2\frac{R_0}{2}}\right)+\gamma_i   \right]\delta({\cal G}_\alpha)-2\frac{\alpha_s}{\pi} {\bf T}_i^2\frac{1}{\alpha-1}\left[ \frac{\Theta({\cal G}_\alpha)}{{\cal G}_\alpha} \right]_+ \ ,
\end{equation}
where $\gamma_i$ is defined in Eq.~\ref{gamma_flav}.  Note the non-trivial dependence of the anomalous dimension on both the
jet radius and the resolution parameter of the angular correlation function.  The anomalous dimension of the measured soft function
for a quark or gluon jet is
\begin{equation}\label{softmeasad}
\gamma_S^{\text{meas}}({\cal G}_\alpha^i) = -\frac{\alpha_s}{\pi}{\bf T}_i^2 \frac{1}{\alpha-1}\left[ \delta({\cal G}_\alpha)\log\frac{\mu^2\tan^{2(\alpha-1)}\frac{R}{2}}{\omega_i^2}-2\left( \frac{\Theta({\cal G}_\alpha)}{{\cal G}_\alpha}\right)_+ \right] \ .
\end{equation}
As mentioned earlier, jet angularity is not factorizable for $a=1$ and here we see that the anomalous dimensions of the
angular correlation jet and soft functions become meaningless for $\alpha=1$, signaling a breakdown of factorization.  For the angular
correlation function, we are most interested in $\alpha=2$, so we will not consider this issue further here.

Summing over the measured jet and soft function anomalous dimensions, we find
\begin{equation}
\sum_{i\in\text{meas}}\left(  \gamma_{J_i}({\cal G}_\alpha^i;\mu)+\gamma_S^{\text{meas}}({\cal G}_\alpha^i;\mu)   \right)=\left(\frac{\alpha_s}{\pi}\sum_{i\in \text{meas}} {\bf T}_i^2 \log\frac{\mu^2}{\omega_i^2\tan^2\frac{R_0}{2}}+\sum_{i\in\text{meas}}\gamma_i \right)\delta({\cal G}_\alpha)
\end{equation}
Note that there is a non-trivial cancellation of the angular correlation function resolution parameter $R$ between
the jet and soft functions.  This contribution exactly cancels that from the hard and unmeasured jet and soft functions in 
Eq.~\ref{unmeas_anom}, consistent with the factorization requirement.

\subsection{\label{resum}Resummation and Averaging}

To proceed with the resummation to NLL of the jet and soft functions, we will make a few observations.  First, as mentioned earlier,
because we are ultimately interested in the average angular structure function, we can ignore factors in the resummed cross section
that are independent of ${\cal G}_\alpha$ or the resolution parameter $R$.  Thus, we will not discuss nor resum the hard function nor
the unmeasured jet and soft functions.  Also, we will only consider a single measured jet in an event.  This prevents a study of inter-jet 
correlations of the angular correlation function, but for this paper we are most interested in the intra-jet dynamics.  Anyway, 
the existence of factorization of jet observables essentially trivializes correlations between jets since it implies that correlations can only come from the 
soft function.  From these observations, we only need to resum the measured jet and soft functions of a single jet.

With these considerations, we will need to compute the convolution between the measured jet and soft functions:
\begin{equation}
\frac{d\sigma}{d{\cal G}_\alpha}\propto\int d{\cal G}'_\alpha \  J({\cal G}_\alpha-{\cal G}'_\alpha;\mu_J,\mu) S({\cal G}'_\alpha;\mu_S,\mu) \ ,
\end{equation}
where $\mu_J$ and $\mu_S$ are the jet and soft scales respectively.  
We refer the reader to \cite{Ellis:2010rwa} for the details of generic NLL-level resummation.  Here, we will use the results collected there
appropriate for the angular correlation function.  
The resummed differential cross section for the angular correlation function of a single measured jet at NLL is
\begin{equation}\label{resum_cs}
\frac{d\sigma}{d{\cal G}_\alpha}\propto \left(\frac{\mu_J}{\omega}  \right)^{\alpha\omega_J}  \left(\frac{\mu_S\tan^{\alpha-1}\frac{R}{2}}{\omega}  \right)^{\omega_S} \left[ 1+f_J({\cal G}_\alpha)+f_S({\cal G}_\alpha)  \right]\frac{e^{K_J+K_S+\gamma_E(\omega_J+\omega_S)}}{\Gamma(-\omega_J-\omega_S)}\left[ \frac{1}{{\cal G}_\alpha^{1+\omega_S+\omega_J}}  \right]_+ \ .
\end{equation}
$\omega$ is the $-$ component of the jet's momentum and $\gamma_E$ is the Euler-Mascheroni constant.
The functions $\omega_J$, $\omega_S$, $K_J$, $K_S$, $f_J$ and $f_S$ are written in detail in Appendix~\ref{resum_app}. They
depend on the jet and soft scales and the factorization scale $\mu$.  The jet and soft scales will be in general sensitive to the  
value of ${\cal G}_\alpha$ and the resolution parameter $R$.  At very small values of ${\cal G}_\alpha$, the
resummed distribution can become negative and in general will need to be matched onto a non-perturbative shape function
in that region.  We do not attempt to correct the shape at very small ${\cal G}_a$ and instead just set the cross section to zero where it
would otherwise be negative.

The average angular correlation function can then be computed from the cross section in Eq.~\ref{resum_cs} by integrating over
${\cal G}_\alpha$:
\begin{eqnarray}
\langle {\cal G}_\alpha(R) \rangle &=& \int_0^{{\cal G}_\alpha^{\max}} d{\cal G}_\alpha \ \frac{d\sigma}{d{\cal G}_\alpha}   {\cal G}_\alpha \nonumber\\
&\propto &   \int_0^{{\cal G}_\alpha^{\max}} d{\cal G}_\alpha   \left(\frac{\mu_J}{\omega}  \right)^{\alpha\omega_J}  \left(\frac{\mu_S\tan^{\alpha-1}\frac{R}{2}}{\omega}  \right)^{\omega_S} \frac{e^{K_J+K_S+\gamma_E(\omega_J+\omega_S)}}{\Gamma(-\omega_J-\omega_S)} \frac{\left[ 1+f_J+f_S \right]}{{\cal G}_\alpha^{\omega_S+\omega_J}} \ ,  
\end{eqnarray}
where ${\cal G}_\alpha^{\max} = \frac{\tan^\alpha\frac{R}{2}}{4}$ is the maximum value of the angular correlation function for a jet with two 
constituents.  We choose the scales $\mu_J$ and $\mu_S$ so as to eliminate the logarithms that remain in the resummed
distribution.  The choice of these scales can be seen from the form of the $f_J$ and $f_S$ terms as given
in the appendix.  We find
\begin{equation}\label{fact_scales1}
\mu_J = \omega {\cal G}_\alpha^{1/\alpha} \ , \qquad \mu_S = \frac{\omega {\cal G}_\alpha}{\tan^{\alpha-1}\frac{R}{2}} \ . 
\end{equation}
With this choice of scales, the average angular correlation function simplifies:
\begin{equation}
\langle {\cal G}_\alpha(R) \rangle \propto \int_0^{{\cal G}_\alpha^{\max}} d{\cal G}_\alpha  \  \frac{e^{K_J+K_S+\gamma_E(\omega_J+\omega_S)}}{\Gamma(-\omega_J-\omega_S)} \left[ 1+f_J+f_S \right] \ .
\end{equation}
Note, however, that there is non-trivial dependence on ${\cal G}_\alpha$ in the functions $\omega_J$, $\omega_S$, $K_J$ and $K_S$.
Finally, to determine $\langle \Delta {\cal G}_\alpha \rangle$, we compute
\begin{equation}
\langle \Delta {\cal G}_\alpha \rangle = \frac{d\log \langle {\cal G}_\alpha \rangle}{d\log R} \ .
\end{equation}
Plots of the average angular structure function as computed in SCET and compared to Monte Carlo and NLO corrections will be presented in the following 
sections.

While resummation of the angular correlation function is necessary for an accurate description of the singular regions of phase space,
it is not obvious how important resummation is for the average angular correlation function.  Resummation of the distribution $d\sigma/d{\cal G}_\alpha$
tames the singularity at small values of ${\cal G}_\alpha$ and produces a peak.  The resummation contribution to the distribution is
most important near the peak while fixed-order contributions are most important in the tail, at large values of ${\cal G}_\alpha$.  However,
the average angular correlation function is sensitive to both the resummed and fixed-order contributions.  To get a sense of the importance
of the resummed contribution, we can compare the location of the peak in $d\sigma/d{\cal G}_\alpha$ to the maximum ${\cal G}_\alpha$ value possible
for a jet with two constituents.  If the ratio of the location of the peak to the maximum value of ${\cal G}_\alpha$ is small, resummation effects
are minimal while if that ratio is large, then the resummed contribution dominates.  


 \begin{figure}
\centering
    \includegraphics[width=8.0cm]{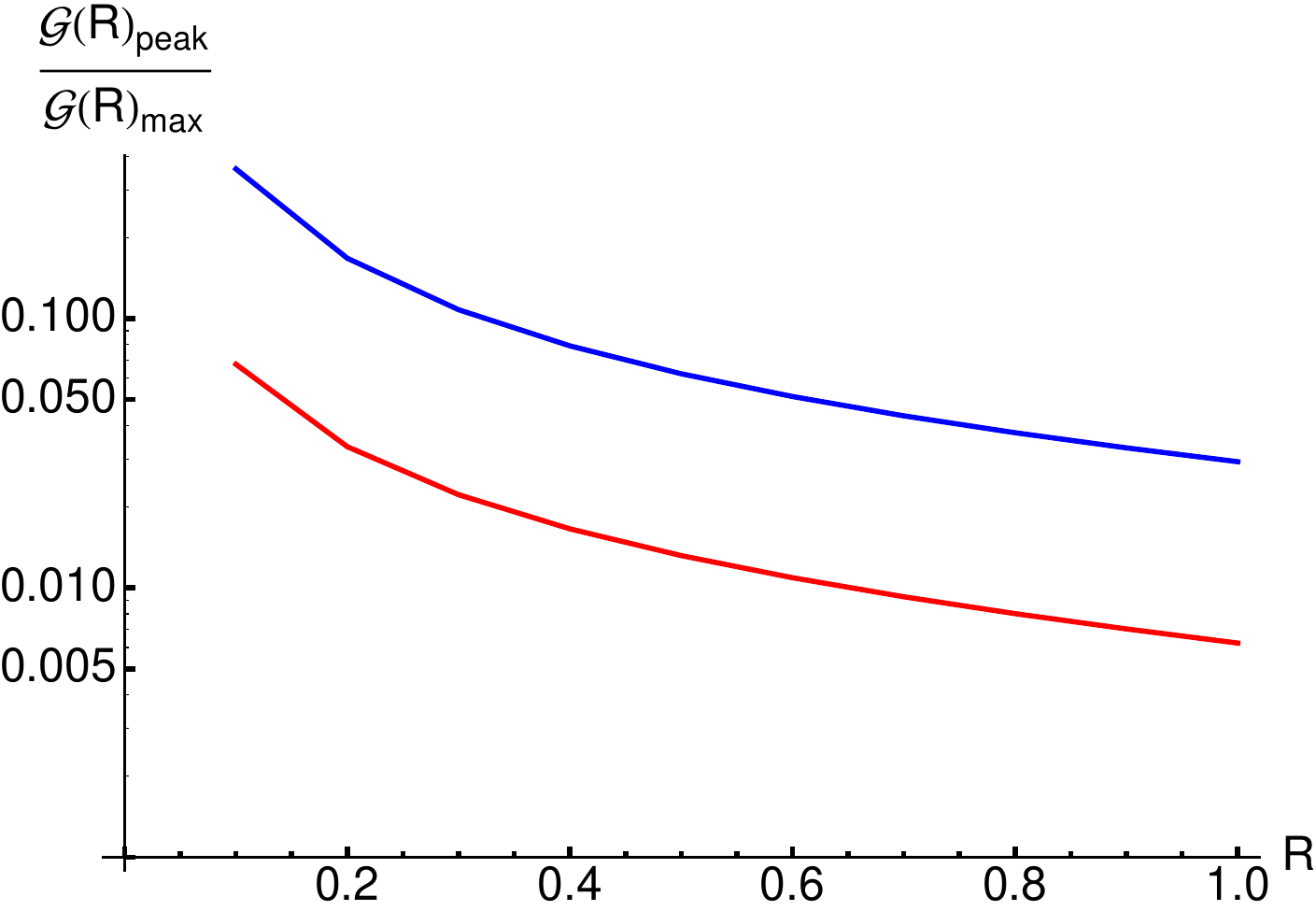}\label{fig:peak_rat_comp}
\caption{
Plots of the ratio between the location of the peak in $\frac{d\sigma}{d\cal{G}_\alpha}$ to the 
maximum value of $\cal{G}_\alpha$ over a range in $R$.  For illustration, $\alpha=2$
and the red (blue) curve is quark (gluon) jets.  The jet radius is $R_0=1.0$ and we have
set the hard, jet and soft scales as in Eq.~\ref{fact_scales1}.
}
\label{comp_plots_peakrat}
\end{figure}


The comparison of the location of the peak in $\frac{d\sigma}{d{\cal G}_\alpha}$ to the maximum value of ${\cal G}_{\alpha}$ is shown in 
Fig.~\ref{comp_plots_peakrat}.  Here, we have set $\alpha=2$ for illustration and the plot shows how the location of the peak
relative to the maximum value changes as the resolution parameter $R$ decreases.  When $R=R_0$, the angular correlation function
is just the jet mass and the ratio is relatively small for both quark and gluon jets.  However, as $R$ is decreased from $R_0$, the ratio increases,
reflecting the greater importance of the resummed contribution with respect to fixed-order corrections.  Thus, we expect that the fixed-order
contribution to the average angular correlation and structure functions is largest at $R\sim R_0$ while the resummed contribution becomes
more important at smaller $R$.  This will be discussed later when comparing the resummed calculation to Monte Carlo and NLO calculation
of the average angular structure function.

\subsubsection{Lowest-Order Expansion}

Before continuing, it is illuminating to expand the angular correlation function to lowest order in the coupling $\alpha_s$.  To do this, we
will need to expand Eq.~\ref{resum_cs} to ${\cal O}(\alpha_s)$.  The form of all of the functions in Eq.~\ref{resum_cs} are given in
Appendix~\ref{resum_app} and, in particular, the expansions of the Gamma, harmonic number and polygamma functions are needed.
The necessary expansions are given in the appendix.  To leading order in $\alpha_s$, we find
\begin{equation}\label{xs_exp}
\frac{d\sigma}{d{\cal G}_\alpha}\propto \frac{\alpha_s(\mu)}{2\pi} {\bf T}_i^2 \left[ 4\log\tan\frac{R}{2}-\frac{4}{\alpha}\log{\cal G}_\alpha   - \frac{1}{\alpha}\left( c_i + \log\frac{\tan^2\frac{R}{2}}{\tan^2\frac{R_0}{2}}  \right)  \right] \frac{1}{{\cal G}_\alpha} + {\cal O}(\alpha_s^2) \ ,
\end{equation}
where the factor $c_i$ depends on the flavor of the jet:
\begin{equation}\label{ci_eq}
c_q = \frac{3}{2} \ , \qquad c_g = \frac{\beta_0}{2 C_A} \ .
\end{equation}
To compute this, we have set the jet and soft scales so as to minimize the logarithms that appear in the cross section as defined in 
Eq.~\ref{fact_scales1}.  In this expression, note that the non-cusp piece of the anomalous dimension of the measured jet functions appears in the term in
parentheses.

From this expression for the cross section differential in the angular correlation function, we integrate over ${\cal G}_\alpha$ to compute the
average angular correlation function.  To ${\cal O}(\alpha_s)$ we find
\begin{eqnarray}\label{ave_asf_lo}
\langle {\cal G}_\alpha \rangle &\equiv& \int_0^{\frac{\tan^a\frac{R}{2}}{4}} d{\cal G}_\alpha \frac{d\sigma}{d{\cal G}_\alpha} {\cal G}_\alpha  \nonumber \\
& \propto & \int_0^{\frac{\tan^\alpha\frac{R}{2}}{4}} \ d{\cal G}_\alpha \frac{\alpha_s(\mu)}{2\pi} {\bf T}_i^2 \left[ 4\log\tan\frac{R}{2}-\frac{4}{\alpha}\log{\cal G}_\alpha   - \frac{1}{\alpha}\left( c_i + \log\frac{\tan^2\frac{R}{2}}{\tan^2\frac{R_0}{2}}  \right)  \right] \frac{1}{{\cal G}_\alpha} {\cal G}_\alpha \nonumber \\
&=& \frac{\alpha_s(\mu)}{2\pi} {\bf T}_i^2\frac{\tan^\alpha\frac{R}{2}}{\alpha}\left(  1 + \log 4 - \frac{c_i}{4} - \frac{1}{4} \log \frac{\tan^2\frac{R}{2}}{\tan^2\frac{R_0}{2}}  \right) + {\cal O}(\alpha_s^2) \ .
\end{eqnarray}
Any overall factor independent of $R$ does not affect the average angular structure function because 
\begin{equation}
\langle \Delta{\cal G}_\alpha(R) \rangle \equiv \frac{d\log\langle {\cal G}_\alpha\rangle}{d\log R} = \frac{R}{\langle {\cal G}_\alpha\rangle}\frac{d\langle{\cal G}_\alpha\rangle}{dR} \ .
\end{equation}
To lowest order, the average angular structure function is independent of $\alpha_s$ and only dependent on the color of the jet through the $c_i$ term.
Eq.~\ref{ave_asf_lo} results in the average angular structure function of
\begin{equation}\label{aveasf_exp}
\langle \Delta {\cal G}_\alpha(R) \rangle = \frac{R}{\sin R}\left(  \alpha- \frac{2}{4+4\log 4 - \left(c_i + \log\frac{\tan^2 \frac{R}{2}}{\tan^2\frac{R_0}{2}}\right)}   \right) +  {\cal O}\left(\alpha_s(\mu)\right) \ .
\end{equation}

Eq.~\ref{aveasf_exp} contains much of the physics that we expect affects the form of the angular structure function.  
The na\"ive expectation for $\langle \Delta {\cal G}_\alpha \rangle$ is $\langle \Delta {\cal G}_\alpha\rangle \simeq \alpha$.  Eq.~\ref{aveasf_exp}
contains an ${\cal O}(1)$ correction to this result that is negative.  This was interpreted
in \cite{Jankowiak:2012na} as an effect due to the running coupling.  However, here, this is probably not the source of this effect because
even for fixed coupling the negative term exists.  This is instead probably due to SCET itself because only collinear and soft emissions
are included with respect to full QCD.  Including all terms in the resummed result should decrease the average angular structure function further
due to both the running coupling and because an arbitrary number of soft and collinear emissions are considered.

Also, note that the term $c_i$ is larger for quarks than for gluons with sufficiently many flavors of quarks:
\begin{equation}
c_q = \frac{3}{2} \geq \frac{11}{6} -\frac{N_f}{9} = c_g \ , 
\end{equation}
for $N_f\geq 3$.  This implies that, for sufficiently many flavors, $\langle \Delta {\cal G}_\alpha\rangle_g > \langle \Delta {\cal G}_\alpha\rangle_q$,
an observation that was also made in \cite{Jankowiak:2012na}.  There, this was attributed to the fact that gluons have more color
than quarks and so radiate more at larger angles, effectively decreasing the strength of the collinear singularity with respect to quarks.
We expect that the resummation magnifies the distinction between quarks and gluons.

Another interesting observation to be made about the form of the angular structure function is that, to this order, it is Lorentz invariant.  We
then expect that all jets, regardless of energy (so long as it is above the hadronization scale of QCD), 
have an angular structure function that deviates only slightly from the form in 
Eq.~\ref{aveasf_exp}.  In particular, note that Eq.~\ref{aveasf_exp} is the infinite jet energy limit of the (all-orders) angular structure function.
The contribution of higher orders to the angular structure function would contain prefactors of $\alpha_s(\mu)$ which would vanish
as $\mu\to \infty$.  If we ignore the finite $R$ terms from the expansion of sine and tangent, $\langle \Delta {\cal G}_\alpha(R)\rangle$ is
very flat, signifying very near scale invariance over a large dynamical range $R$.  Flatness is only broken by a term that goes like 
$1/\log\frac{R}{R_0}$ which is only important at very small $R/R_0$.

It is accurate, then, to represent the angular structure function in the form (again, ignoring the finite $R$ terms from sine and tangent)
\begin{equation}\label{a_shift}
\langle \Delta {\cal G}_\alpha(R) \rangle \simeq \alpha - \gamma_{\text{ASF}} \ ,
\end{equation}
where $\gamma_{\text{ASF}}$ might be called the anomalous dimension of a QCD jet and is independent of $\alpha$.  This anomalous
dimension is a robust quantity that is intrinsic to the flavor of the jet and properties of QCD.  Measuring this property
of the angular structure function in data would be very interesting.  It is important to note, however, for all of the above comments, ${\cal O}(\alpha_s)$
contributions to the average angular structure function have been ignored.  These are expected to be comparable in size to the second term in
Eq.~\ref{aveasf_exp}.  Note in particular that NNLO contributions can be just as, or even more, important than the contributions from resummation.
Indeed, for jets with three constituents, it was computed in \cite{Jankowiak:2012na} that the effect at this order is to increase the average angular structure function.

\subsection{\label{hadron_sec}Non-Perturbative Physics Effects}

In addition to the perturbative physics contribution to the angular structure function, we would also like to understand the effects
from non-perturbative physics.  For jets produced in an $e^+e^-$ collider, the dominant non-perturbative effect is from hadronization.
A simple physical argument can be used to determine how hadronization affects the angular structure function.  The partons created
from the parton shower will be connected to one another by color strings which stretch across the event.  After the termination of the parton shower
at an energy scale of about 1 GeV, these color strings are allowed to break to create a quark-antiquark pair if it is energetically favorable.  
This string breaking continues until all particles are connected by strings with sufficiently low tension and are then associated into hadrons.
In the process of breaking the strings and creating quark pairs, the number of particles that are created at small angles with respect to one
another increases from that which was created in the perturbative parton shower.  Thus, hadronization increases particle
production at small angles, effectively increasing the strength of the collinear singularity and decreasing the value of
the average angular structure function.

The effect of hadronization decreasing the average angular structure function can also be quantitatively studied.  Note that the angular
correlation function is just the (squared) mass of a jet from constituents that are separated by angular scale $R$ or less.  Dasgupta, Magnea
and Salam \cite{Dasgupta:2007wa} 
studied the effect of non-perturbative physics on the transverse momentum and mass distributions of jets at hadron colliders.  For the mass, they found
that the leading correction due to hadronization is
\begin{equation}
\langle \delta M^2 \rangle \sim C R_0 + {\cal O}(R_0^3) \ ,
\end{equation}
where $C$ is independent of $R_0$, the jet radius.  For the angular correlation function, we expect that the effect of hadronization would also 
result in a correction proportional to $R$, the resolution parameter of the angular correlation function.  We can write
\begin{equation}
\langle{\cal G}_\alpha(R)\rangle \simeq C_{\text{pert}} R^\alpha+C_{\text{non-pert}} R \ ,
\end{equation}
where $C_{\text{pert}}$ is the perturbative contribution to the angular correlation function and $C_{\text{non-pert}}$ is the non-perturbative
contribution.  The average angular structure function that follows from this is
\begin{equation}
\langle \Delta {\cal G}_\alpha(R) \rangle = \frac{\alpha C_{\text{pert}} R^\alpha+C_{\text{non-pert}} R}{C_{\text{pert}} R^\alpha+C_{\text{non-pert}} R} < \alpha \ ,
\end{equation}
where the inequality follows when $\alpha>1$.  Note that the perturbative angular structure function is approximately $\alpha$ and so, indeed, 
hadronization effects decrease the value of the angular structure function.

The argument presented here and in \cite{Dasgupta:2007wa} relies on the one-gluon approximation to determine the effect of hadronization.  
Universality of the hadronization and power correction effects was argued with the one-gluon approximation in 
refs.~\cite{Dokshitzer:1995zt,Dokshitzer:1997ew} and demonstrated for event shapes in SCET in refs.~\cite{Lee:2006nr,Lee:2006fn}.  The arguments
in refs.~\cite{Lee:2006nr,Lee:2006fn} relied on the boost invariance of the soft function for back-to-back jets.  How the argument might
extend to an arbitrary number of jets in arbitrary directions is unclear as the boost invariance is, at least na\"ively, broken.  We will not discuss how this
might be extended, but we note that, because of the qualitative and quantitive arguments from the one-gluon approximation, we expect that
the universality holds in SCET.

\subsection{\label{lhc_asf}The Angular Correlation Function at the LHC}

Finally, we will discuss how the results obtained here for the SCET resummation might be extended to the LHC, to processes initiated by
$pp$ collisions.  For an observable ${\cal O}$ that factorizes at hadron colliders, the cross section can be written in the schematic form 
\cite{Bauer:2002nz,Bauer:2008jx,Stewart:2009yx}
\begin{equation}\label{LHC_fact}
\frac{d\sigma}{d{\cal O}}=H(\mu)\times C_{ab}B_a(\mu) B_b(\mu)\otimes \left[ \prod_{n_i} J_{n_i}({\cal O};\mu) \right]\otimes S({\cal O};\mu)
\end{equation}
The beam functions $B_i$ encode the properties of the initial parton $i$ and the matrix $C_{ab}$ weights the colliding partons by the appropriate
cross section.  Indices $a$ and $b$ are implicitly summed over.
  In this case, the flavor of the jet functions depends on the flavor of the initial colliding partons which affects the admixture of quark and
gluon jets that contribute to ${\cal O}$.  Note also that the soft function includes contributions from radiation from initial state partons.
Therefore, while not necessarily manifest in Eq.~\ref{LHC_fact}, the beam functions implicitly affect the jet and soft functions.

Nevertheless, we expect that the angular correlation function has nice factorization properties at hadron colliders.  With the goal of computing the 
average angular structure function, we can again ignore anything in the factorization of the angular correlation function that is independent of ${\cal G}_a$
or the resolution parameter $R$:
\begin{equation}
\frac{d\sigma}{d{\cal G}_{\alpha i}}\propto  C_{ab}B_a(\mu) B_b(\mu)\otimes J_{n_i}({\cal G}_{\alpha i};\mu) \otimes S_{n_a,n_b;n_1\cdots n_N}({\cal G}_{\alpha i};\mu) \ ,
\end{equation}
where we have chosen to measure ${\cal G}_\alpha$ in jet $i$ in an event with $N$ jets.  Dependence on the beam functions has been retained, however.  This is
because, for a given set of jets $1,\ldots,N$, different initial states contribute to the cross section with different weights.  Thus, the beam function contribution
to the factorization, $C_{ab}B_a(\mu) B_b(\mu)$, is actually not an overall constant factor and so must be included.  Note also that the color of the colliding partons
affects the radiation included in the soft function.  The beam functions are universal and so can be computed once and for all.  While this is not a rigorous 
proof of factorization of the angular correlation function, many of the results obtained in the $e^+e^-$ collider context should be able to be recycled 
for the hadron collider case.  This deserves significant future study.


 \begin{figure}[t]
\centering
\subfigure[ \ SCET vs Pythia8]
{
    \includegraphics[width=7.5cm]{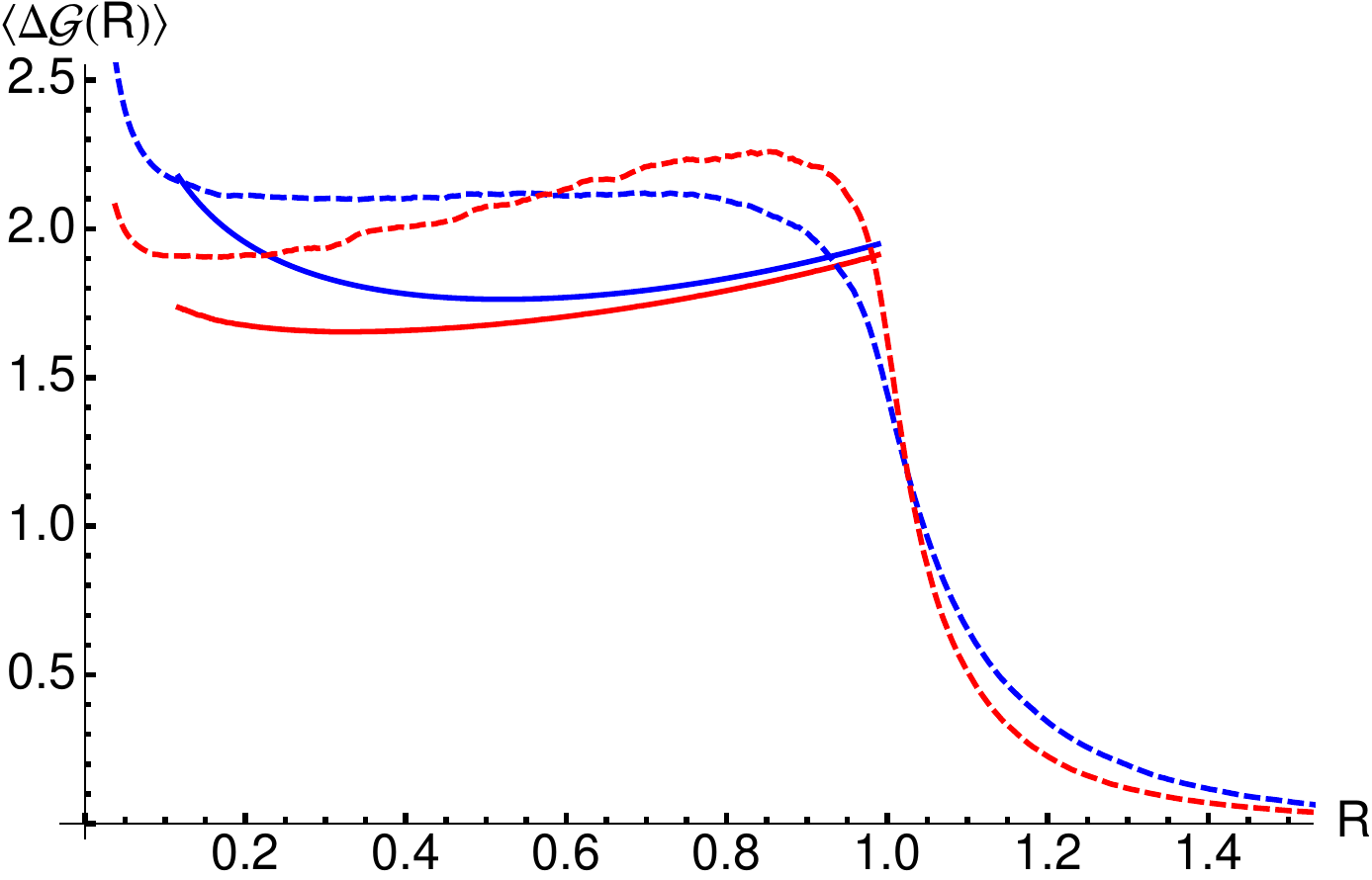}\label{fig:scet_py_comp}
}
\hspace{0cm}
\subfigure[ \ Hard scale variation]
{
    \includegraphics[width=7.5cm]{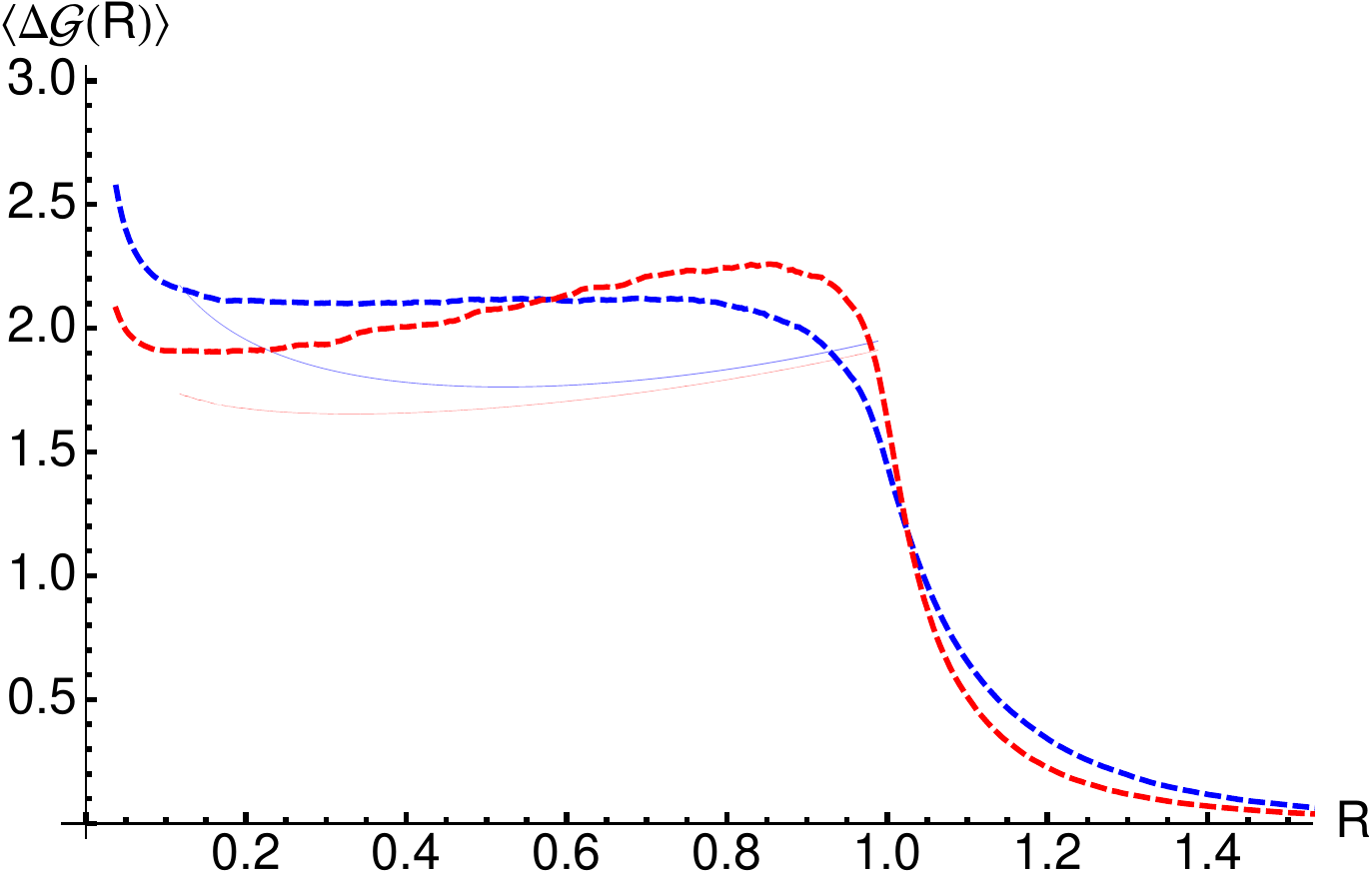}\label{fig:scet_py_comp_h}
}
\hspace{0cm}
\subfigure[ \ Jet scale variation]
{
    \includegraphics[width=7.5cm]{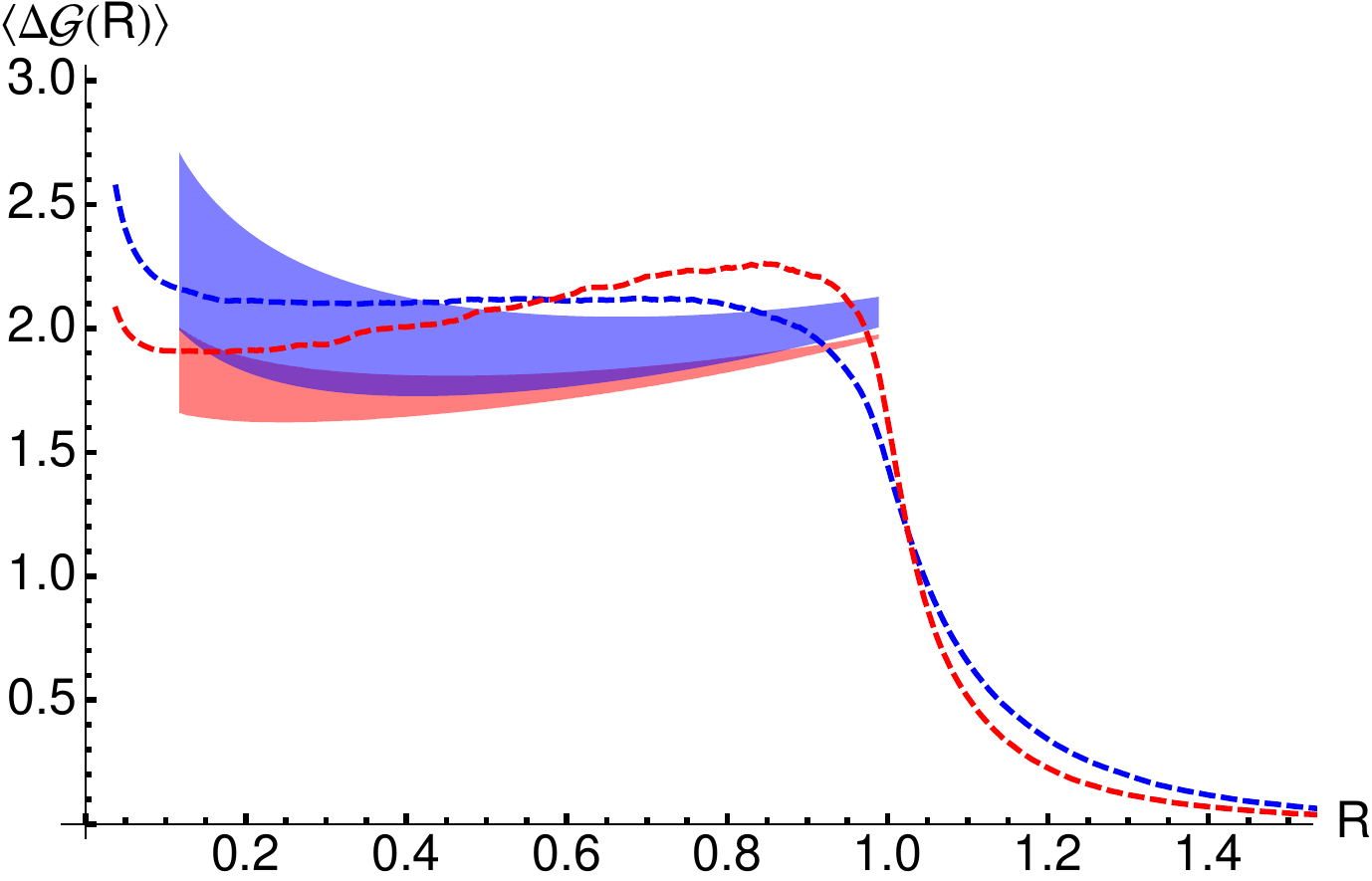}\label{fig:scet_py_comp_j}
}
\hspace{0cm}
\subfigure[ \ Soft scale variation]
{
    \includegraphics[width=7.5cm]{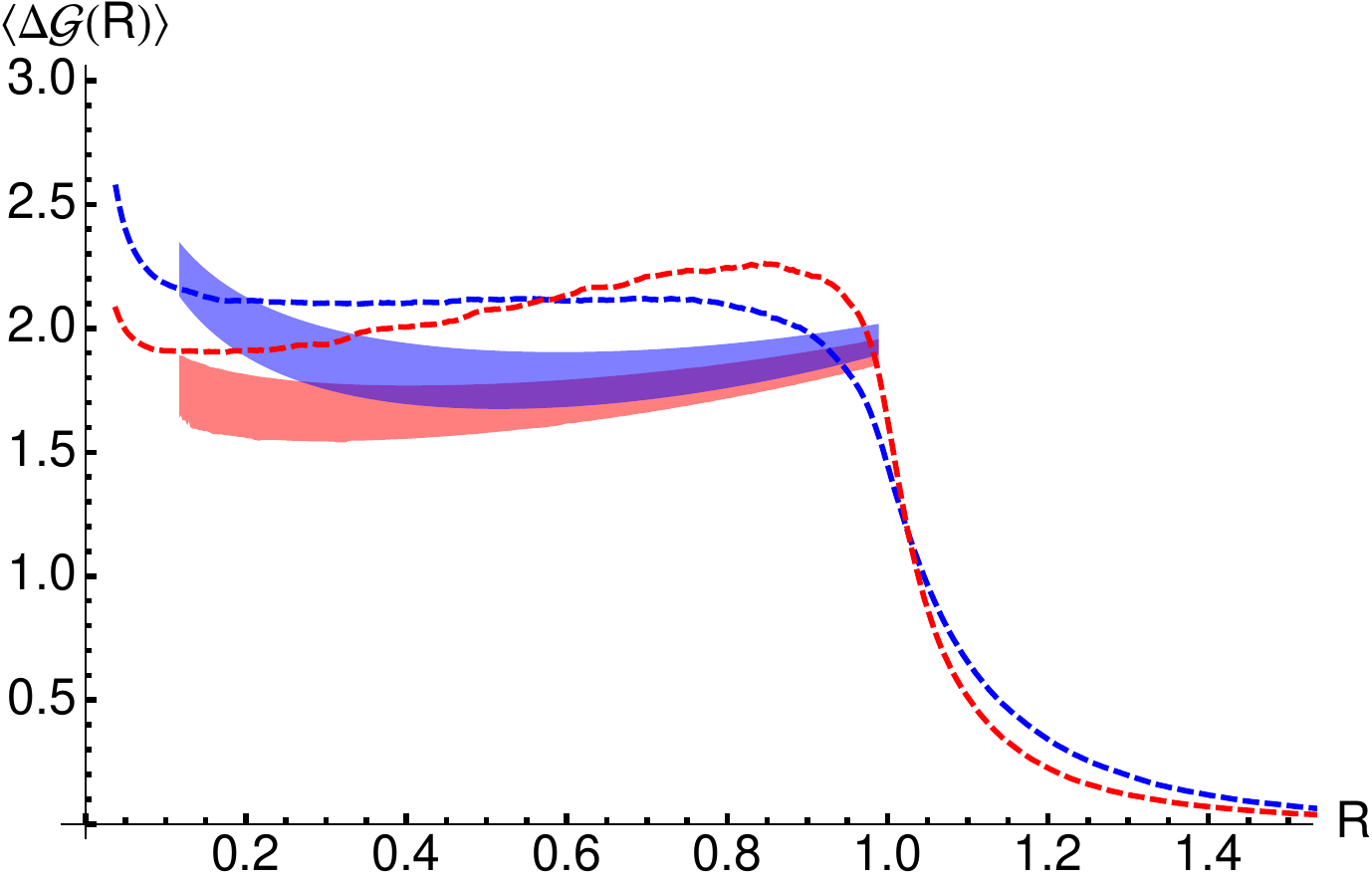}\label{fig:scet_py_comp_s}
}
\caption{Plots of the average angular structure function for quark (red) and gluon (blue) jets.  
Fig.~\ref{fig:scet_py_comp} compares the curves from SCET resummation (solid) to
anti-$k_T$ jets from Pythia8 (dashed).  The Pythia8 curves were computed from 3 jet final states in which all jets had
equal energy.  Figs.~\ref{fig:scet_py_comp_h}, \ref{fig:scet_py_comp_j} and \ref{fig:scet_py_comp_s} compare the 
Pythia8 curves to SCET bands in which the hard, jet and soft scales have been varied by a factor of 2.
To make these curves, the jet radius has been set to be $R_0 = 1.0$ and the energy of the jets is 300 GeV.
}
\label{comp_plots_py_scet}
\end{figure}

\section{\label{FOsec}Comparison to Fixed-Order Calculation}

In this and the following section, we will focus most of our attention on the (proper) angular correlation function with $\alpha=2$:
\begin{equation}\label{acf_2}
{\cal G}_2(R)\equiv {\cal G}(R)=\frac{1}{2E_J^2}\sum_{i\neq j} E_i E_j \sin\theta_{ij}\tan\frac{\theta_{ij}}{2}\Theta(R-\theta_{ij}) \ .
\end{equation}
To evaluate this jet observable in SCET, we must choose the hard, jet and soft scales.  For many of the comparison plots we choose the following scales:
\begin{equation}\label{scale_var}
\mu_H = \omega \ , \qquad \mu_J = \omega {\cal G}^{1/2} \ , \qquad \mu_S = \frac{\omega {\cal G}}{\tan\frac{R}{2}} \ .
\end{equation}
These choices of scales minimize logarithms that appear in the resummed distribution.  However, it is important to understand the dependence
of the result on the choice of these scales and so we will also present plots in which the scales are varied by the standard factors of $2$ and $1/2$.
The evaluation of the average angular structure function from the SCET cross section is done numerically.  Note that for consistency of the factorization
the jet scale $\mu_J$ must be larger than the soft scale $\mu_S$ which is the requirement that
\begin{equation}
\tan\frac{R}{2}\gg {\cal G}^{1/2} \ .
\end{equation}
To maintain this separation, the resolution parameter cannot be too small; we will only consider $R \gtrsim 0.1$.  For smaller values of $R$, 
logarithms of $R$ become large and must be resummed, which is beyond the scope of this paper. 

The average angular structure function as computed in SCET is plotted in Fig.~\ref{comp_plots_py_scet}
where the curves for quark and gluon jets are compared to the output of Pythia8.  The Pythia8 curves will be discussed in the next section.
Fig.~\ref{fig:scet_py_comp} compares the quark and gluon curves with the hard, jet and soft scales set to their values in Eq.~\ref{scale_var}.
The observations from the previous section are apparent with the quark average angular structure function less than the gluon average angular
structure function and both slightly less than 2.  The scale variations of these curves are shown in Figs.~\ref{fig:scet_py_comp_h}, \ref{fig:scet_py_comp_j}
and \ref{fig:scet_py_comp_s}.  Note that in particular there is relatively wide range over which the angular structure function varies when the jet
and soft scales are changed by a factor of 2.  

Because the average angular correlation function is defined by integrating over the entire range of ${\cal G}_\alpha$, its value and shape is
sensitive to radiation in all regions of phase space.  Resummation is necessary for an accurate description of the physics in the singular
regions of phase space while higher fixed-order contributions are necessary for a good description in the non-singular regions of phase space.
A proper treatment of resummation and fixed-order involves consistently matching the two contributions so that the resulting distribution is
accurate order by order in $\alpha_s$ over the entire phase space.  This matching is a non-trivial procedure and, instead, we will
just focus on the contribution from higher fixed-order matrix elements.  This will give us a sense, at least, for how fixed order and resummation
affect the average angular structure function.

To do this, we use NLOJet++ v.~4.1.3 \cite{Nagy:1998bb,Nagy:2003tz}, based on the dipole subtraction method of \cite{Catani:1996vz},
to compute the average angular structure function to NLO in $e^+e^-$ collisions.  NLOJet++ can compute matrix elements to NLO for up to 
4 final state partons (and, at tree level, up to 5 final state partons) 
and so, by demanding jet requirements, produces jets with very few constituents
in them.  This results in very inefficient calculation of cross sections.  Also, the public version of NLOJet++ does not record flavor information of 
partons so the identity of quark and gluon jets cannot be easily determined. 
Further, it is not enough that the cross section differential in the angular correlation function at fixed $R$ is smooth for the average
angular structure function to be smooth.  The distributions must also be smooth over $R$ so that the derivative that defines the average
angular structure function is well-behaved.
 To assuage these issues, in this section, we will define an event-wide angular correlation 
function, where the sum in Eq.~\ref{acf_mod} runs over all particles in the event.

The event-wide angular correlation function is defined over all particles in the event with no jet algorithm cut.
In the limit that there are three final state particles this reduces
precisely to the angular correlation function of the hardest jet, extending up to an $R$ of about the radius of the hardest jet.  
The angular correlation function will only include the contribution from the two closest
partons because the third parton must be very far away in angle.  
This argument doesn't hold at higher orders, but for those cases we expect that the
event-wide definition will be an average over the angular correlation functions of quark and gluon jets.


 \begin{figure}
\centering
    \includegraphics[width=8.0cm]{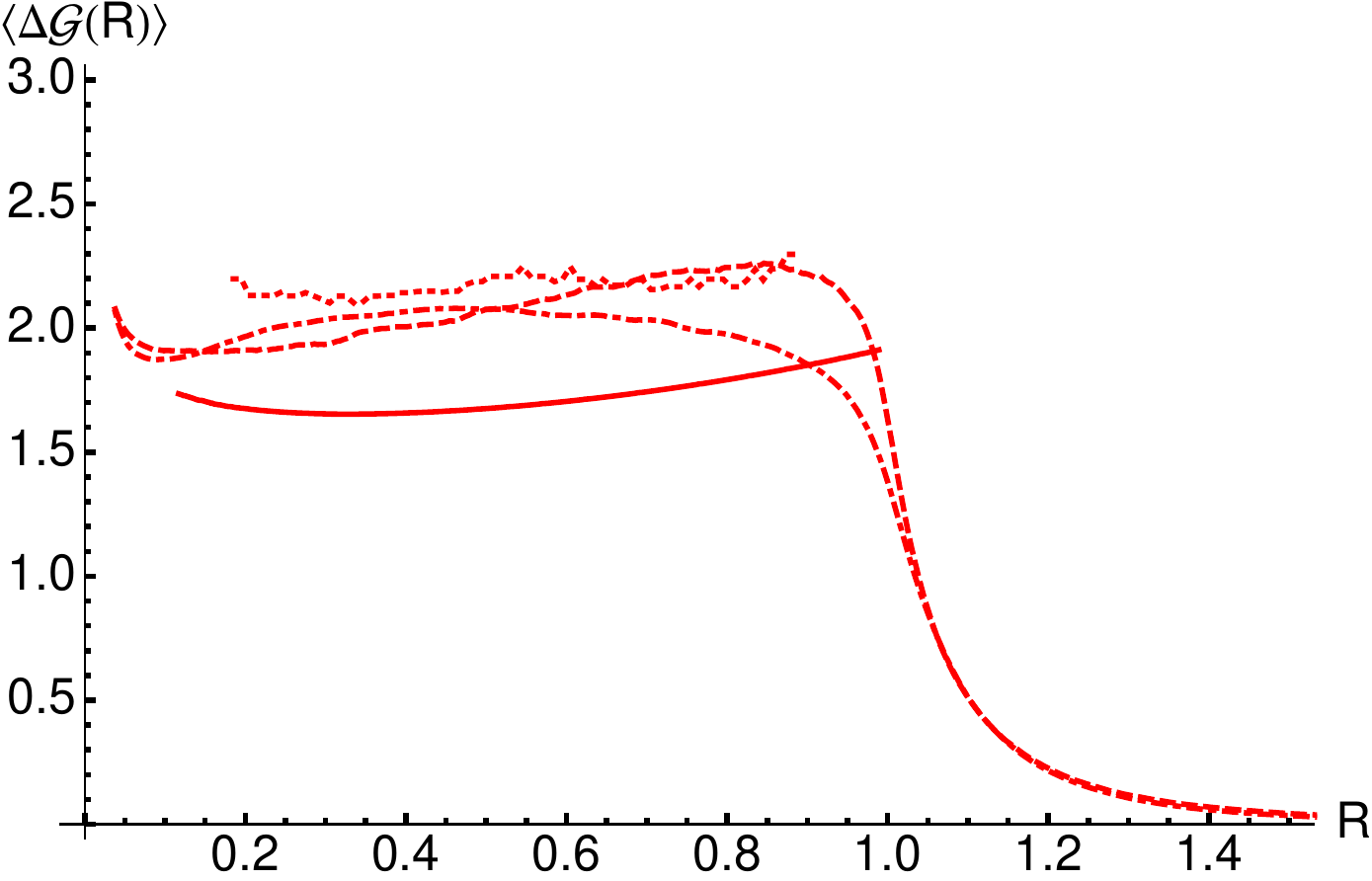}\label{fig:nlo_scet_py}
\caption{Comparison of NLOJet++ calculation of the event-wide average angular structure function in 3 jet final states
to NLO (dotted) to SCET NLL resummation of the average angular structure function for quark jets (solid).  Two curves
from Pythia8 are shown: the dashed curve is the average angular structure function for quark jets from $e^+e^-\to$ 3 jets and the 
dot-dashed curve is the average angular structure function from $e^+e^-\to$ 2 jets.
}
\label{comp_plots_nlo}
\end{figure}


We present the calculation of the average angular structure function from NLOJet++ for three final state partons to NLO in Fig.~\ref{comp_plots_nlo}.
The center of mass energy is taken to be 600 GeV in $e^+e^-$ collisions.
At a $e^+e^-$ collider, most of the time, the hardest jet will contain a quark and a radiated gluon so we compare the output of NLOJet++ to 
NLL resummation results for quark jets.  Fig.~\ref{comp_plots_nlo} also contains two curves of quark jets from Pythia8
which will be discussed in the next section.
  The NLO calculation of the angular structure function is approximately flat and greater than 2 which we interpret
as an effective weakening of the collinear singularity due to the presence of wide-angle radiation.  
The fact that the NLO result is slightly larger that 2 was anticipated in \cite{Jankowiak:2012na} where it was shown that a jet with three constituents
should have an average angular structure function larger than 2 by a term proportinal to $\alpha_s$.
Matching the calculations from NLL  and NLO  would 
produce a curve that interpolates between the NLL result at small $R$ and the NLO result at large $R$.  

To generate the NLOJet++ curve, about one trillion events were processed over about 1 CPU year.  Even with this many events, the 
average angular structure function from NLOJet++ is still quite noisy.  However, the noise can be reduced by averaging the angular structure function
over a small range in $R$ at each point.  This was done for the curve in Fig.~\ref{comp_plots_nlo}.  Computing the average angular
structure function in 4 jet final states to NLO was attempted in the same CPU time as the 3 jet results.  However, the resulting curves were 
much too noisy to be used.   To produce curves at higher orders using NLOJet++ probably requires centuries of CPU time for distributions to converge.
However, other programs such as BlackHat \cite{Berger:2008sj} might be better-suited to higher multiplicity final states at NLO.  Work in this direction is ongoing.

\section{\label{MC}Parton Shower Monte Carlo Comparison}

In this section, we compare our calculation of the average angular structure function from SCET to the output
of Monte Carlo event generator and parton shower.  Through the Sudakov factor which dictates the probability that no branchings occur
between two scales of an evolution variable, the parton shower resums logarithms of the evolution variable
 that arise from soft and collinear emissions.  Monte Carlo generators create
fully exclusive events and so the process of resummation of the logarithms is distinct from that in SCET, for example, and examining
the differences is interesting.  The parameter that defines the evolution in the parton shower is also (relatively) arbitrary and 
different choices of the evolution variable lead to different emphases on soft or collinear splittings.
In addition, hadronization and other non-perturbative physics is described by phenomenological models
which can be used to understand the size and effect of power-suppressed contributions to observables.  All of these points and their
effects will explored in this section.

For most of the Monte Carlo comparison, we first generated tree-level events for the process $e^+ e^- \to q\bar{q}g$ using MadGraph5 v.~1.4.5 
\cite{Alwall:2011uj} at center-of-mass energy of 900 GeV.  These partons are required to each have equal energy $E = 300$ GeV so that they are 
well-separated and factorization-breaking terms in the soft function are minimized.  These events were then showered using the 
$p_T$-ordered shower of Pythia8 v.~8.162 \cite{Sjostrand:2007gs}.  All default settings of Pythia8 were used except for turning hadronization
on and off to study the difference.  In most plots hadronization in Pythia8 has been turned off.  To study the effect of using different
evolution variables in the parton shower we shower the MadGraph events with VINCIA v.~1.0.28 \cite{Giele:2011cb}.  From the showered events,
jets were found with the FastJet v.~3.0.2 \cite{Cacciari:2011ma} implementation of the anti-$k_T$ algorithm \cite{Cacciari:2008gp}. 
 We choose the jet radius to be $R_0 = 1.0$.
The three hardest jets are required to have energy between 250 and 350 GeV and we identify jets as coming from a quark or gluon by demanding that 
the cosine of the angle between the jet axis and the direction of a parton from MadGraph be greater than $0.9$.  

In Fig.~\ref{comp_plots_py_scet}, we plot the average angular structure function for quark and gluon jets identified in Pythia8
(with no hadronization) and the angular structure
function as computed in SCET.  Note that the average angular structure function as computed in SCET is significantly smaller than that from Pythia8, 
especially at larger $R$.  This difference can be attributed to higher order effects which were shown in the previous section to increase the value of the
average angular structure function.
Fig.~\ref{comp_plots_py_scet} also illustrates the distinction between quark and gluon jets.  For most of the range of $0<R<1$, the average angular
structure function for gluon jets is greater than that for quark jets, reflecting the fact that gluons have more color and radiate more at wider angles
than do quarks.  This effect is present in both the Pythia8 curves and the resummed calculation.  Because the SCET calculation only included
effects from jets with at most two constituents, the curves terminate precisely at the jet radius of $R_0=1.0$.  For these anti-$k_T$ jets in Pythia8, the
edge effects from the jet algorithm are small, extending only over a range of at most $R=0.8$ to $R=1.2$.  Also, we have not plotted the SCET curves
below $R=0.1$, where they begin to deviate substantially from their value at larger $R$.

Fig.~\ref{comp_plots_nlo} compares the average angular structure function from NLOJet++ to quark jets in
SCET and two different curves from Pythia8.  The different Pythia8 curves exhibit the affect of wide angle radiation captured
by the jet on the average angular structure function.  In that figure, the dashed curve is the quark jet average angular structure function
from the Pythia8 sample described above.  The dot-dashed curve is the the average angular structure function from quark jets 
from $e^+e^- \to q\bar{q}$ samples generated and showered in (otherwise default) Pythia8.  The center of mass energy was set to be $600$ GeV
and the jets were required to have energy within 50 GeV of 300 GeV.  Higher order effects are obvious.  The Pythia8
curves agree well with one another at small $R$ up to $R\sim 0.6$ and then diverge at larger $R$.  The jets in the 3 jet sample collect wide-angle
radiation from the neighboring jets which increases the average angular structure function at large $R$.  To fully understand the
rise within an analytic calculation requires matching fixed-order to resummed result.  Fixed-order contributions are responsible for the wide-angle emissions
that increase the average angular structure function because SCET factorization effectively decouples the jets.

\begin{figure}
\centering
\subfigure[ \ Identified gluon jets]
{
    \includegraphics[width=8.0cm]{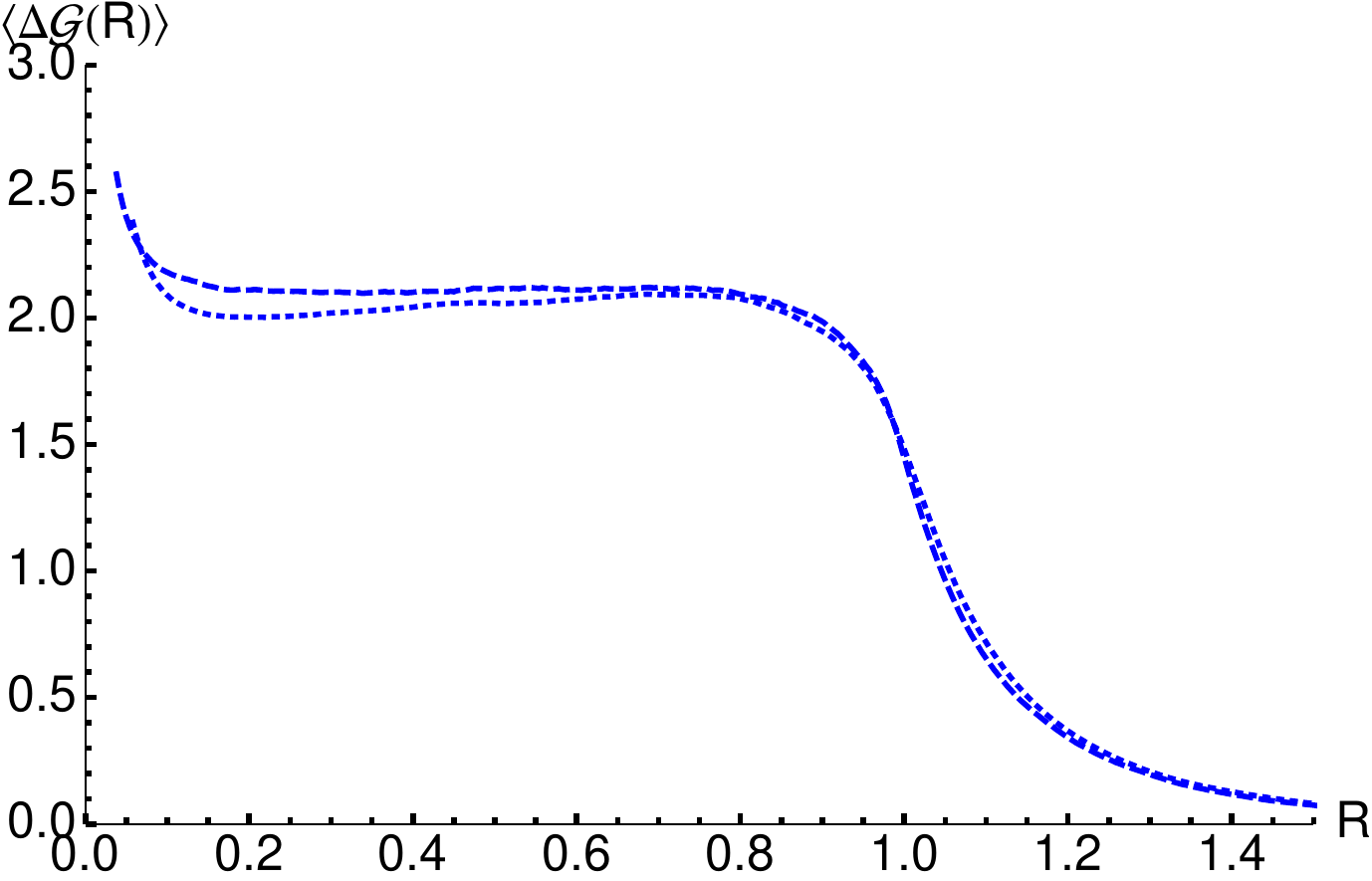}\label{fig:py_had_comp_g}
}
\hspace{-.4cm}
\subfigure[ \ Identified quark jets]
{
    \includegraphics[width=8.0cm]{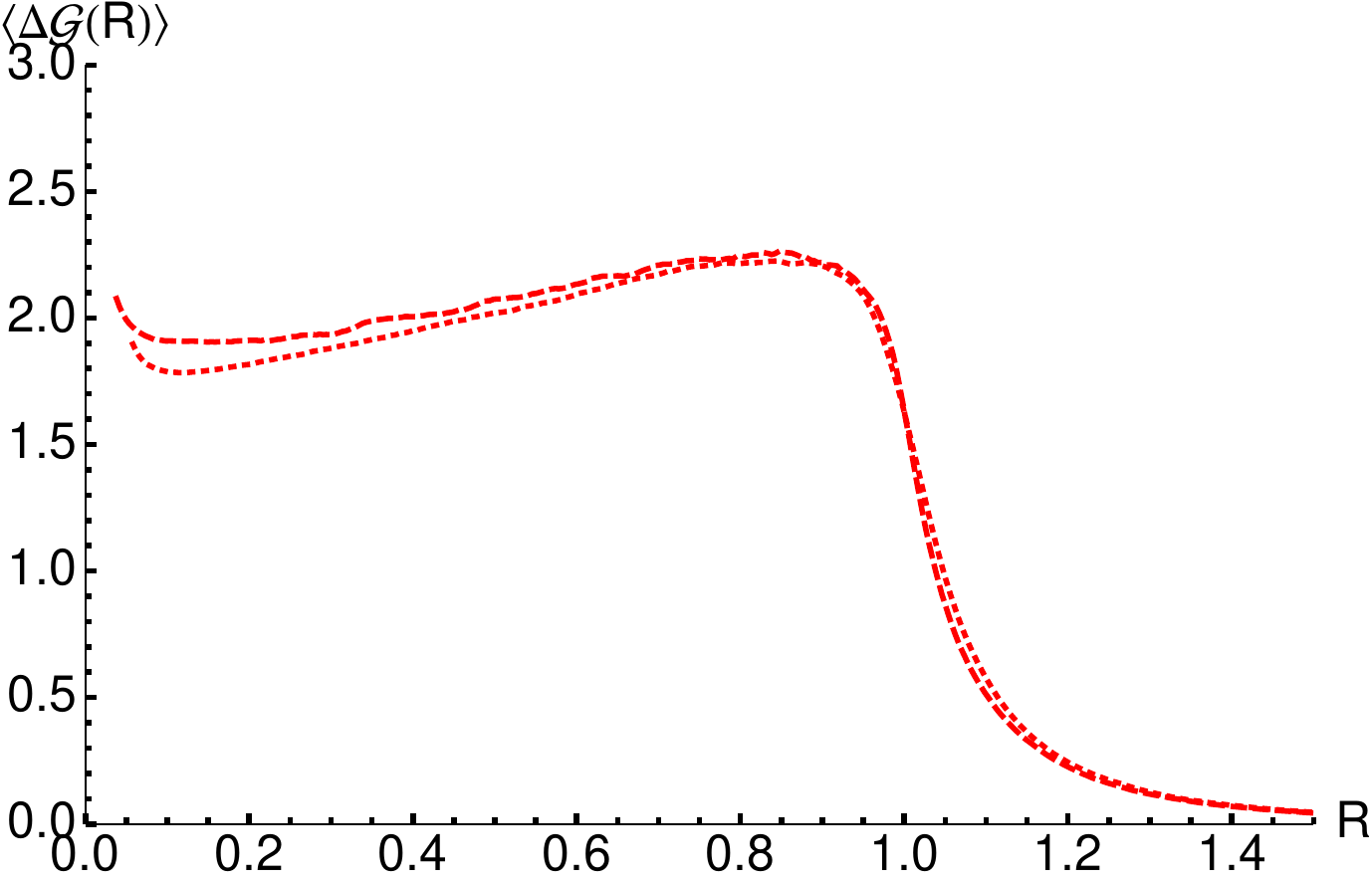}\label{fig:py_had_comp_q}
}
\caption{Comparison of the average angular structure function as computed in Pythia8 with (dotted) and without (dashed) hadronization.
}
\label{comp_plots_had}
\end{figure}

As discussed in Sec.~\ref{hadron_sec}, we expect the effect from non-perturbative physics on the angular structure function to be 
small and relatively well-understood.  In particular, relying on arguments from the one-gluon approximation, we expect that hadronization
increases the strength of the collinear singularity and that this effect is most prominent at small values of $R$.  In Fig.~\ref{comp_plots_had},
we have plotted the average angular structure function for quarks and gluons comparing the curves with hadronization turned on or off.
Indeed, the effect is small but unambiguous: hadronization effectively increases the strength of the collinear singularity.
  As discussed earlier,
extending the arguments from \cite{Lee:2006nr,Lee:2006fn} on the effect of non-perturbative physics would be greatly desired to fully describe
(at least) average behavior of hadronization for events with multiple jets.


\subsection{Monte Carlo Error Estimates}
Finally in this section, we would like to get a handle on the error or uncertainty in the Monte Carlo parton shower in Pythia.  Typically, this 
is done by studying the output of different tunes of the same Monte Carlo program or comparing different Monte Carlo programs altogether.  
In particular, as is relevant for the parton shower,
the evolution variable of the parton shower dictates when and how emissions should occur.  
  Ref.~\cite{Banfi:2010xy} observed differences
in event shape variables as computed in Pythia 6.4 \cite{Sjostrand:2006za} between two tunes; one $p_T$-ordered and the other 
virtuality ordered.  However, these two tunes had other distinctions as well and so purely the effect of the evolution variable is obscured.  Also,
comparing two different Monte Carlos is subtle because the number of differences is typically huge and so isolating effects of single parameters
or choices is very difficult.


 \begin{figure}
\centering
\subfigure[ \ Identified gluon jets]
{
    \includegraphics[width=8.0cm]{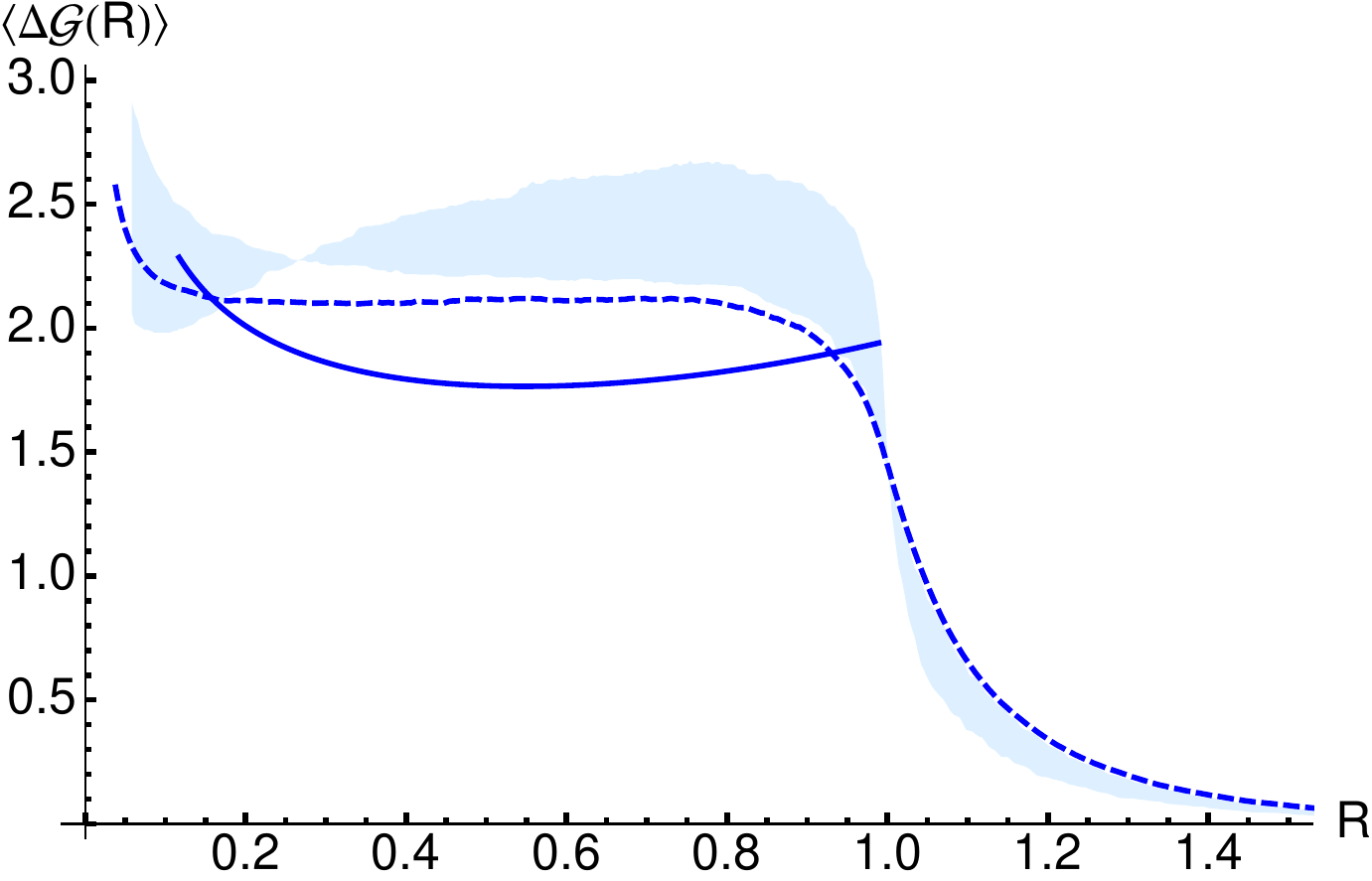}\label{fig:py_vin_scet_comp_g}
}
\hspace{-.4cm}
\subfigure[ \ Identified quark jets]
{
    \includegraphics[width=8.0cm]{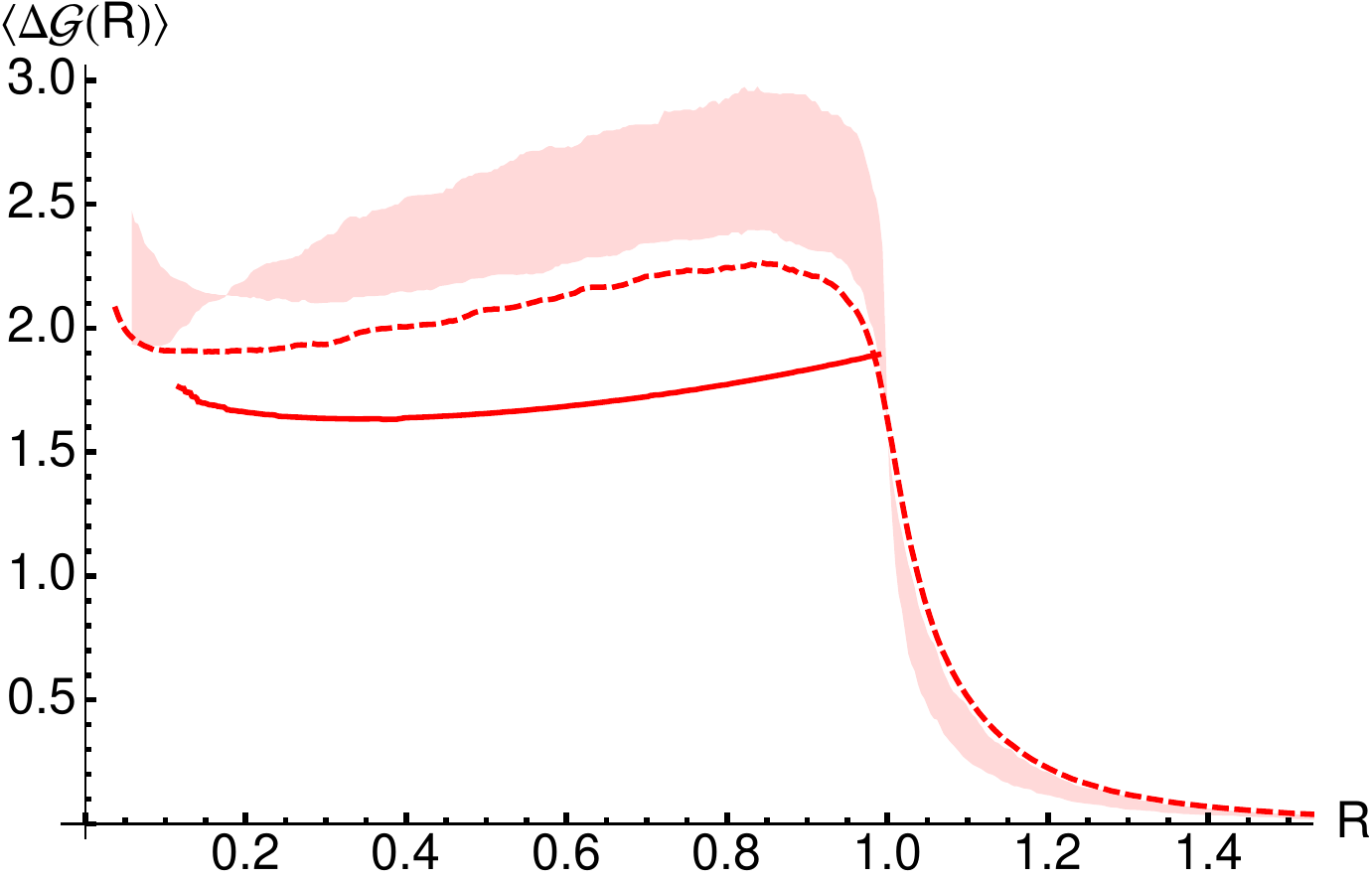}\label{fig:py_vin_scet_comp_q}
}
\caption{Comparison of SCET computation (solid) and Pythia8 (dashed) of average angular structure function to the 
output of VINCIA Monte Carlo parton shower with two different evolution variables: $p_T$ and virtuality.  The shaded region
lies between the curves from VINCIA.
}
\label{comp_plots_v}
\end{figure}


Here, we would like to study the effect of different evolution variables in the parton shower.  The choice of the evolution variable is
only a change of variables in the Sudakov form factor and so must produce the exact same leading-log resummation for any (consistent) choice of
evolution variable.  However, the choice of evolution variable can lead to higher log-order effects through the scale at which $\alpha_s$ is
evaluated or by emphasizing soft over collinear splittings, for example.  To study the differences, we use the VINCIA \cite{Giele:2011cb}
parton shower plug-in for Pythia8 which is based on 2-to-3 splittings as opposed to the standard 1-to-2 splittings as in Pythia and Herwig
\cite{Bahr:2008pv}.  VINCIA includes a flag which allows the user to change only the evolution variable.  For concreteness, we will consider
$p_T$-ordering and virtuality ordering.  

In Fig.~\ref{comp_plots_v} we have plotted the SCET resummation and Pythia8 output for the average angular structure function as well
as a band which extends over the range between the output of the $p_T$-ordered and the virtuality ordered shower in VINCIA.  
The exact same requirements on the jets were made in the VINCIA sample as in the Pythia8 sample as described earlier.  Over most of the
range in $R$, the lower edge of the VINCIA band is set by the $p_T$-ordered shower and upper edge by the virtuality ordered shower.  This is
expected as the $p_T$-ordered shower emphasizes collinear emissions more than the virtuality-ordered shower.  Note also that the band is slightly 
above the output of the $p_T$ ordered shower in Pythia8.  Part of this effect could be due to the default matrix element matching in VINCIA: 
final states with up to 5 partons are matched to tree-level matrix elements.  
Regardless of the details, the effect of changing the evolution variable is large.  Understanding if and how parton showers resum higher order logarithms
with different evolution variables, matching schemes, {\it etc.}, is necessary to understand the source of the differences.

\section{\label{conc}Conclusions}
The average angular correlation and structure functions capture the average scaling properties of QCD jets.  We have presented a calculation
of the angular correlation function to NLL accuracy in SCET and compared this result to the Pythia8 Monte Carlo parton shower and to fixed-order 
results from NLOJet++.  Comparing the resummed SCET result to the fixed-order NLOJet++ result provides a good understanding as to the behavior
of the parton shower result.
However, for a full understanding, matching of the resummed and fixed-order distributions is required.
Much like the jet shape \cite{Ellis:1992qq}, the average angular structure function could be used for tuning of the Monte Carlo.  Because it is a two-point correlation
function, the angular correlation function captures distinct information from the jet shape and so this tuning would be non-trivial.  

There are several directions for extending the study presented here.  First, it would be desirable to compute the angular correlation function in collisions
at the LHC.  It remains an outstanding problem to use SCET to resum logarithms for arbitrary observables in hadron colliders because factorization
of the (colored) initial and final states is highly non-trivial.  However, using the observations from Sec.~\ref{lhc_asf}, the computation of the average angular
structure function at hadron colliders might only require a reinterpretation of the results presented here.
  Recently, NLO results were obtained for $pp\to 4j$ events \cite{Bern:2011ep} from which any IRC-safe
observable could be computed.  In particular, for four final state partons, the hardest jet can contain up to three constituents which would be beyond the
resummed order in the SCET calculation.  
Also, at a hadron collider, underlying event or pile-up produce significant background radiation that can be collected
into a jet.  A procedure to determine the contribution to a jet from these non-perturbative sources is necessary to properly determine jet energy scales and 
to study substructure.  The results presented here could be used to determine the average contribution to a jet using the procedure introduced in
\cite{Jankowiak:2012na}.

For a more accurate prediction of the angular correlation function, matching of NLL and NLO results must be done to have good control of the distribution
over the entire phase space.  Factorization of jet observables allows for a process-independent computation of the NLL resummed result; however, the 
fixed-order calculation is process dependent and must be couched in a particular study.  We showed that the average angular structure function is 
sensitive to wide-angle radiation so matching is vital for accurate predictions.
  NLOJet++ or results like those from \cite{Bern:2011ep}
are promising in their applicability to generic processes.  It is unlikely that QCD jet observables can be reliably computed to NNLL or beyond analytically
because non-global logarithms become important.  Nevertheless, by studying limiting behavior such as in \cite{Feige:2012vc} the effect of these non-global
logarithms might be reduced. 

Finally, as there exist few jet substructure observables that have been (or even can be) computed with analytic methods, it is important to compute those
that are possible.  The calculation of the angular correlation function provides powerful insight into the behavior of QCD and the dynamic properties of
jets.  Though scale-invariance is broken in QCD by a running coupling, jets maintain a fractal, conformal structure to very good approximation
over a wide dynamical range.  

\appendix
\section{\label{jet_func_app}Measured Jet Functions}

Here, we present the finite pieces of the measured jet functions for quark and gluon jets as defined by a $k_T$-type
algorithm for the angular correlation
function.  These functions are composed of contributions from $\delta$-functions and $+$-distributions.  For a function
$g(x)$, we define the $+$-distribution as \cite{Ligeti:2008ac}
\begin{equation}
\left[ g(x)\Theta(x)\right]_+ = g(x)\Theta(x)-\delta(x)\int_0^1 dx'\ g(x') \ ,
\end{equation}
so that
\begin{equation}
\int_0^1 dx \ \left[ g(x)\Theta(x)\right]_+ = 0 \ .
\end{equation}
From this definition, it is straightforward to compute the measured jet functions.  The terms that are infinite in four dimensions were presented
in Sec.~\ref{measjet}.  The terms that are finite in four dimensions are, for a quark jet:
\begin{eqnarray}\label{Qjetfunc_app}
J_{\omega}^{q}({\cal G}_\alpha,\epsilon^0)&=& \frac{\alpha_s C_F}{2\pi}\left\{ \left( \frac{13}{2} - \frac{9\alpha-8}{12(\alpha-1)}\pi^2+\frac{3}{2}\log\frac{\mu^2}{\omega^2\tan^2\frac{R_0}{2}} + \frac{\alpha/2}{\alpha-1}\log^2\frac{\mu^2}{\omega^2}\right.\right.\nonumber\\
&&\left.\left.+\log\frac{\mu^2}{\omega^2}\log\frac{\tan^2\frac{R}{2}}{\tan^2\frac{R_0}{2}} +\frac{1}{2}\log^2 \tan^2\frac{R_0}{2}+\frac{\alpha-1}{2}\log^2\tan^2\frac{R}{2} \right)\delta({\cal G}_\alpha) \right.\nonumber\\
&&\left. + \left[\Theta({\cal G}_\alpha)\Theta({\cal G}_\alpha^{\max}-{\cal G}_\alpha)\left(\frac{4}{\alpha-1}\frac{\log {\cal G}_\alpha}{{\cal G}_\alpha}   - \frac{2}{\alpha-1}\frac{\log\frac{\mu^2}{\omega^2{}\tan^{2(\alpha-1)}\frac{R}{2}}}{{\cal G}_\alpha}\right.\right.\right.\nonumber\\
&&+\left.\left.\left. \frac{4}{\alpha}\frac{1}{{\cal G}_\alpha} \log\frac{1+\sqrt{1-\frac{4{\cal G}_\alpha}{\tan^\alpha\frac{R}{2}}}}{1-\sqrt{1-\frac{4{\cal G}_\alpha}{\tan^\alpha\frac{R}{2}}}} -\frac{3}{\alpha} \frac{\sqrt{1-\frac{4{\cal G}_\alpha}{\tan^\alpha\frac{R}{2}}}}{{\cal G}_\alpha} \right) \right]_+\right\}\ .
\end{eqnarray}
For a gluon jet, the finite terms are:
\begin{eqnarray}\label{Gjetfunc_app}
J_{\omega}^{g}({\cal G}_\alpha,\epsilon^0)&=& \frac{\alpha_s}{2\pi}\left\{ 
\left(   \frac{67}{9}C_A - \frac{23}{9}n_F T_R-\frac{9\alpha-8}{12(\alpha-1)}\pi^2+\frac{\beta_0}{2}\log\frac{\mu^2}{\omega^2\tan^2\frac{R_0}{2}}\right. \right.\nonumber \\
&&\left.+ C_A\frac{\alpha/2}{\alpha-1}\log^2\frac{\mu^2}{\omega^2}+C_A\log\frac{\mu^2}{\omega^2}\log\frac{\tan^2\frac{R}{2}}{\tan^2\frac{R_0}{2}}+\frac{1}{2}\log^2\tan^2\frac{R_0}{2}\right.\nonumber\\
&&\left.\left.+\frac{\alpha-1}{2}\log^2\tan^2\frac{R}{2}\right)\delta({\cal G}_\alpha) +\left[\Theta({\cal G}_\alpha)\Theta({\cal G}_\alpha^{\max}-{\cal G}_\alpha)\left(   - \frac{\beta_0}{\alpha} \frac{\sqrt{1-\frac{4{\cal G}_\alpha}{\tan^\alpha\frac{R}{2}}}}{{\cal G}_\alpha}  \right.\right.\right. \nonumber\\
&& \left. 
+ \frac{4C_A}{\alpha} \frac{1}{{\cal G}_\alpha}\log\frac{1+\sqrt{1-\frac{4{\cal G}_\alpha}{\tan^\alpha\frac{R}{2}}}}{1-\sqrt{1-\frac{4{\cal G}_\alpha}{\tan^\alpha\frac{R}{2}}}}    
+\frac{2C_A-4n_F T_R}{3\alpha\tan^\alpha\frac{R}{2}} \sqrt{1-\frac{4{\cal G}_\alpha}{\tan^\alpha\frac{R}{2}}}\right.\nonumber\\
&&\left.\left.\left.
+ \frac{4C_A}{\alpha-1} \frac{\log {\cal G}_\alpha}{{\cal G}_\alpha}  -\frac{2C_A}{\alpha-1}\frac{\log\frac{\mu^2}{\omega^2 \tan^{2(\alpha-1)\frac{R}{2}}}}{{\cal G}_\alpha} \right)\right]_+
\right\} \ .
\end{eqnarray}
${\cal G}_\alpha^{\max}$ is the largest value that ${\cal G}_\alpha$ can take for a jet with two constituents:
\begin{equation}
{\cal G}_\alpha^{\max} = \frac{\tan^\alpha\frac{R}{2}}{4} \ ,
\end{equation}
where we have taken the leading $\lambda$ dependence for the collinear modes.

\section{\label{resum_app}Resummed Distribution for Angular Correlation Function}
The expression for the resummed cross section is
\begin{equation}\label{resum_cs_app}
\frac{d\sigma}{d{\cal G}_\alpha}\propto \left(\frac{\mu_J}{\omega}  \right)^{\alpha\omega_J}  \left(\frac{\mu_S\tan^{\alpha-1}\frac{R}{2}}{\omega}  \right)^{\omega_S} \left[ 1+f_J({\cal G}_\alpha)+f_S({\cal G}_\alpha)  \right]\frac{e^{K_J+K_S+\gamma_E(\omega_J+\omega_S)}}{\Gamma(-\omega_J-\omega_S)}\left[ \frac{1}{{\cal G}_\alpha^{1+\omega_S+\omega_J}}  \right]_+ \ .
\end{equation}
$\omega_{J}$, $\omega_S$, $K_J$ and $K_S$ result from the resummation of the individual jet and soft functions \cite{Korchemsky:1993uz,Becher:2006mr,Balzereit:1998yf,Neubert:2005nt,Fleming:2007xt}.
  The functions
$\omega_J$ and $\omega_K$ are defined by $\omega_J \equiv \omega_F(\mu,\mu_J)$ and $\omega_S \equiv - \omega_F(\mu,\mu_S)$
where
\begin{equation}
\omega_{F}(\mu,\mu_0) = -\frac{4{\bf T}_i^2}{(\alpha-1)\beta_0} \left[ \log r + \left( \frac{\Gamma^1_{\text{cusp}}}{\Gamma^0_{\text{cusp}}} -\frac{\beta_1}{\beta_0} \right) \frac{\alpha(\mu_0)}{4\pi}\left(r-1\right) \right] \ .
\end{equation}
$\beta_0$ is the coefficient of the one-loop $\beta$-function as defined in Eq.~\ref{beta_one} and $\beta_1$ is the two-loop
coeffcient:
\begin{equation}
\beta_1 = \frac{34}{3}C_A^2-\frac{10}{3}C_A N_f-2C_FN_f \ .
\end{equation}
$r$ is the ratio of the strong coupling at two scales:
\begin{equation}
r=\frac{\alpha_s(\mu)}{\alpha_s(\mu_0)} \ ,
\end{equation}
and the energy dependence of the strong coupling is given by the two-loop expression
\begin{equation}
\frac{1}{\alpha_s(\mu)} = \frac{1}{\alpha_s(Q)}+\frac{\beta_0}{2\pi}\log\left(\frac{\mu}{Q}\right)+\frac{\beta_1}{4\pi \beta_0}\log\left[  1+\frac{\beta_0}{2\pi}\alpha_s(Q)\log\left( \frac{\mu}{Q} \right) \right] \ ,
\end{equation}
for $\alpha_s$ evaluated at the two scales $\mu$ and $Q$.  The terms $\Gamma_{\text{cusp}}^0$ and $\Gamma_{\text{cusp}}^1$ 
are the one- and two-loop coefficients of the cusp anomalous dimension.  Their ratio is given by \cite{Korchemsky:1987wg} 
\begin{equation}
\frac{\Gamma^1_{\text{cusp}}}{\Gamma^0_{\text{cusp}}} = \left( \frac{67}{9} - \frac{\pi^2}{3}  \right)C_A - \frac{10}{9}N_f \ .
\end{equation}

The function $K_J$ is given by
\begin{eqnarray}
K_J(\mu,\mu_J) =  -\frac{\gamma^0_i}{2\beta_0}\log r - \frac{8\pi \alpha {\bf T}_i^2 }{(\alpha-1)\beta_0^2} &&\left[ \frac{r-1-r\log r}{\alpha_s(\mu)} \right. \nonumber\\
&&\left.+ \left( \frac{\Gamma^1_{\text{cusp}}}{\Gamma^0_{\text{cusp}}} -\frac{\beta_1}{\beta_0} \right) \frac{1-r+\log r}{4\pi} + \frac{\beta_1}{8\pi\beta_0}\log^2 r \right] \ , \ 
\end{eqnarray}
with $\gamma^0_i$ defined as
\begin{equation}
\gamma_i^0 =  4{\bf T}_i^2 \left( \log\frac{\tan^2\frac{R}{2}}{\tan^2\frac{R_0}{2}}+c_i  \right)  \ ,
\end{equation}
where $c_i$ is defined in Eq.~\ref{ci_eq}.  $K_S$ is defined similarly
\begin{equation}
K_S(\mu,\mu_S) = \frac{8\pi  {\bf T}_i^2 }{(\alpha-1)\beta_0^2} \left[ \frac{r-1-r\log r}{\alpha_s(\mu)}+ \left( \frac{\Gamma^1_{\text{cusp}}}{\Gamma^0_{\text{cusp}}} -\frac{\beta_1}{\beta_0} \right) \frac{1-r+\log r}{4\pi} + \frac{\beta_1}{8\pi\beta_0}\log^2 r \right] \ .
\end{equation}

The functions $f_J$ and $f_S$ are generated by the convolution of the jet and soft functions.  Accurate to NLL, they are
\begin{eqnarray}\label{fj_app}
f_J({\cal G}_\alpha;\mu,\mu_J) &=& \frac{\alpha_s(\mu_J){\bf T}_i^2}{2\pi}\Theta\left( {\cal G}_\alpha^{\max} - {\cal G}_\alpha \right) \left\{\frac{2\alpha}{\alpha-1}\log^2\frac{\mu_J}{\omega {\cal G}_\alpha^{1/\alpha}} + \log^2\tan^2\frac{R_0}{2} \right.\nonumber\\
&&+\  (\alpha-1)\log^2\tan^2\frac{R}{2} - 2c_i \log \tan\frac{R_0}{2}+\frac{1}{\alpha-1}\frac{2}{\alpha}\left[ \frac{\pi^2}{6} - \psi^{(1)}(-\omega_J-\omega_S) \right]   \nonumber\\
&&+\left[ c_i + \log\frac{\tan^2\frac{R}{2}}{\tan^2\frac{R_0}{2}} +\frac{2}{\alpha-1}H(-1-\omega_J-\omega_S) \right] \nonumber\\
&&\left. \ \ \ \ \ \ \  \times  \left[ 2\log\frac{\mu_J}{\omega {\cal G}_\alpha^{1/\alpha}} + \frac{1}{\alpha}H(-1-\omega_J-\omega_S) \right] \right\} \ ,
\end{eqnarray}
and
\begin{eqnarray}\label{fs_app}
f_S({\cal G}_\alpha;\mu,\mu_S) &=&  -\frac{\alpha_s(\mu_S)}{\pi}\frac{{\bf T}_i^2}{\alpha-1} \left[  \left(  \log \frac{\mu_S \tan^{\alpha-1}\frac{R}{2}}{\omega {\cal G}_\alpha} + H(-1-\omega_J-\omega_S) \right)^2 \right.\nonumber\\
&&  \left. \ \ \ \ \ \ \ \ \ + \ \frac{\pi^2}{6} - \psi^{(1)}(-\omega_J-\omega_S)\right]\ .
\end{eqnarray}
$H(x)$ is the harmonic number function defined by
\begin{equation}
H(x) = \int_0^1\frac{1-t^x}{1-t}\ dt \ ,
\end{equation}
and $\psi^{(1)}(x)$ is the trigamma function
\begin{equation}
\psi^{(1)}(x) = \int^{\infty}_0 \frac{te^{-xt}}{1-e^{-t}} \ dt \ .
\end{equation}
Note that the logarithms in these functions can be minimized by choosing
\begin{equation}\label{scales_app}
\mu_J = \omega {\cal G}_\alpha^{1/\alpha} \ , \qquad \mu_S = \frac{\omega {\cal G}_\alpha}{\tan^{\alpha-1}\frac{R}{2}} \ .
\end{equation}

For expansion of the resummed distribution, the following relations are needed:
\begin{eqnarray}
H(-1-\epsilon)^2 - \psi^{(1)}(-\epsilon) &=&  -\frac{\pi^2}{2} + {\cal O}(\epsilon) \ ,\\
\frac{H(-1-\epsilon)}{\Gamma(-\epsilon)} &=& -1+\gamma_E\epsilon+{\cal O}(\epsilon^2) \ .
\end{eqnarray}

\begin{acknowledgments}
A.~L.~thanks Jon Walsh for extensive correspondence on SCET and the computation of jet and soft functions as well as detailed comments on a draft
of this paper.  
A.~L.~also thanks Gavin Salam for pointing out the discussion of a similar angular correlation function moment in the literature.
A.~L.~also thanks Michael Peskin for his insisting that this work be done and helpful comments, especially in comparing calculation to Monte Carlo.  
This work is supported by the US Department of Energy under contract DE--AC02--76SF00515 and partial support by the U.S. National Science Foundation, grant NSF--PHY--0969510, the LHC Theory Initiative, Jonathan Bagger, PI.
\end{acknowledgments}

\bibliography{ACF_SCET}

\end{document}